\documentclass[graybox]{svmult}
\usepackage{amsmath}
\usepackage{mathrsfs}
\usepackage{amssymb}

\usepackage{mathptmx}       
\usepackage{helvet}         
\usepackage{courier}        
\usepackage{type1cm}        
\usepackage{makeidx}         
\usepackage{graphicx}        
\usepackage{multicol}        
\usepackage[bottom]{footmisc}

\makeindex             

\begin{document}

\title*{Simulation of heat transport in low-dimensional oscillator lattices}
\tableofcontents

\author{Lei Wang, Nianbei Li, Peter H\"anggi}
\institute{Lei Wang \at Department of Physics and Beijing Key Laboratory of Opto-electronic Functional Materials and Micro-nano Devices,
Renmin University of China, Beijing 100872, People's Republic of China,\\
\email{phywanglei@ruc.edu.cn}
\and
Nianbei Li \at Center for Phononics and Thermal Energy Science, School of Physics Science and Engineering, Tongji University, Shanghai 200092, People's Republic of China,\\
\email{nbli@tongji.edu.cn}
\and
Peter H\"anggi \at Institute of Physics, University of Augsburg, Augsburg D-86135, Germany,\\
\email{Peter.Hanggi@physik.uni-augsburg.de}}

\maketitle

\abstract
{The  study of heat transport in low-dimensional oscillator lattices presents a formidable challenge. Theoretical efforts have been made trying to reveal the underlying mechanism of diversified heat transport behaviors. In lack of a unified rigorous treatment, approximate theories often may embody controversial predictions. It is therefore of ultimate importance that one can rely on numerical simulations in the investigation of heat transfer processes in low-dimensional lattices. The simulation of heat transport using the non-equilibrium heat bath method and the Green-Kubo method will be  introduced. It is found that one-dimensional (1D), two-dimensional (2D) and three-dimensional (3D) momentum-conserving nonlinear lattices display power-law divergent, logarithmic divergent and constant thermal conductivities, respectively. Next, a novel diffusion method is also introduced. The heat diffusion theory connects the energy diffusion and heat conduction in a straightforward manner. This enables one to use the diffusion method to investigate the objective of heat transport. In addition, it contains fundamental information about the heat transport process which cannot readily be gathered otherwise.}

\section{Simulation of heat transport with non-equilibrium heat bath method and equilibrium Green-Kubo method}\label{sec:wang}

We start out by considering numerically heat transport in typical momentum-conserving nonlinear lattices, from 1D to 3D.
The numerical simulations are performed with two different methods: the non-equilibrium heat bath method and the celebrated equilibrium Green-Kubo method.
Our major focus will be on  the length dependence of thermal conductivities where the asymptotic behavior towards the thermodynamic limit is of prime interest. As a result, numerical simulations are usually taken on lattices employing very large up to even huge system sizes. Therefore, in order to get a compromise between better numerical accuracy and acceptable computational cost, a 5-th order Runge-Kutta algorithm~\cite{butcher1993} is applied for the simulations of the dissipative systems in the former case, while an embedded Runge-Kutta-Nystrom algorithm of orders 8(6)~\cite{IMAJNA.7.423,IMAJNA.11.297}
is applied for the simulations of the conservative Hamiltonian systems in the latter case.

\subsection{Power-law divergent thermal conductivity in 1D momentum-conserving nonlinear lattices}\label{sec:1d}

Heat conduction induced by a small temperature gradient is expected to satisfy the Fourier's law in the stationary regime:
\begin{equation} \label{eq:kne}
j=-\kappa \nabla T,
\end{equation}
where $j$ denotes the steady state heat flux, $\nabla T$ denotes the small temperature gradient, and $\kappa$ denotes the thermal conductivity. In practical numerical simulations, the temperature difference $\Delta T$ is usually fixed for convenience. In this setup, for a system with length $L$, the steady state heat flux $j$ should be inversely proportional to $L$: $j=-\kappa \Delta T/L$, if Fourier's law is obeyed and $\kappa$ is a constant. However, for many 1D momentum-conserving lattices~\cite{PhysRep.377.1,AdvPhys.57.457}, it is numerically found that $j$ decays as $L^{-1+\alpha}$ with a positive $\alpha$. This finding indicates that the thermal conductivity $\kappa$ is length dependent and diverges with $L$ as $\kappa\propto L^{\alpha}$ in the thermodynamical limit $L\rightarrow\infty$. The Fourier's law is broken and the heat conduction is called anomalous.

For this anomalous heat conduction, transport theories from different approaches make different predictions for the divergency exponent $\alpha$~\cite{AdvPhys.57.457}. The renormalization group theory~\cite{PhysRevLett.89.200601} for 1D fluids predicts a universal value of $\alpha=1/3$ and it is claimed that the thermal conductivity of oscillator chains including the Fermi-Pasta-Ulam (FPU) lattices should diverge in this universal way~\cite{PhysRevE.73.061202}. Early Mode-Coupling Theories (MCT) predict one universal value of $\alpha=2/5$ for all 1D
FPU lattices~\cite{PhysRep.377.1}, while another MCT taking the transverse motion into account predicts $\alpha=1/3$ ~\cite{PhysRevLett.92.074302,PhysRevE.70.021204}. Later, a self-consistent MCT proposes that there should be two universality classes instead of one.
It states that models with asymmetric interaction potentials are characterized by a divergency exponent $\alpha=1/3$,
while models with symmetric potentials are characterized by a larger value of $\alpha=1/2$~\cite{PhysRevE.73.060201,JStatMech.2007.P02007,EurPhysJSpecialTopics.146.21}.
The value of $\alpha=1/3$ is also predicted by calculating the relaxation rates of phonons~\cite{PhysRevE.77.011113}. -- For an intriguing  discussion of the physical existence  of anharmonic phonons and its interrelation between a phonon mean free path
and its associated mean phonon relaxation time we refer the interested readers to recent work \cite{Liu_Liu2014}.

A theory based on Peierls-Boltzmann equation is applied for the FPU-$\beta$ lattice, and $\alpha=2/5$ is predicted\cite{PhysRevE.68.056124}.
In Ref.~\cite{CommPureApplMath.61.1753}, energy current correlation function is studied for the FPU-$\beta$ lattice and $\alpha=2/5$ is found with small nonlinearity approximation. Similar to the discrepancy among theoretical predictions, numerical results are also not convergent. For example,
 an early numerical study suggests $\alpha=2/5$~\cite{PhysRevLett.78.1896}, while some recent studies support $\alpha=1/3$ ~\cite{PhysRevLett.98.184301,PhysRevLett.100.199401,PhysRevLett.100.199402}.

In this section, we numerically study heat conduction in typical 1D nonlinear lattices with the following Hamiltonian
\begin{equation}
H =\sum_i [\frac{p^2_i}2+V(u_i-u_{i-1})],
\label{eq:Ham}
\end{equation}
where $p_i$ and $u_i$ denotes the momentum and displacement from equilibrium position for $i$-th particle. For convenience, dimensionless units are applied and the mass of all particles can be set as unity. The interaction potential energy between particles $i$ and $i-1$ is $V_i \equiv V(u_i-u_{i-1})$. The interaction force is correspondingly obtained as $f_i=-{\partial V_i}/{\partial u_i}$.
The local energy belongs to the particle $i$ is defined here as $E_i=\frac{\dot p^2_i}2+\frac12 (V_i+V_{i+1})$,
i.e., neighboring particles share their interaction potential energy equally.
The instantaneous local heat flux is then defined as $j_i \equiv \frac12(\dot u_i+\dot u_{i+1})f_{i+1}$ and the total heat flux is defined as $J(t)\equiv\sum_i j_i(t)$.

The interaction potential takes the general FPU form as
$V(u)=\frac{1}{2}k_2 u^2+\frac{1}{3} k_3 u^3+\frac{1}{4}k_4 u^4$.
The following three types of lattices will be studied, i.e.,
(1) the FPU-$\alpha\beta$ lattices with $k_2=k_4=1, k_3=1$ (in short as FPU-$\alpha1\beta$ lattice); and
 $k_2=k_4=1, k_3=2$ (in short as FPU-$\alpha2\beta$ lattice);
(2) the FPU-$\beta$ lattice with $k_2=k_4=1$, $k_3=0$; and
(3) the purely quartic or the qFPU-$\beta$ lattice with $k_2=k_3=0$, $k_4=1$.
The interactions in the FPU-$\alpha\beta$ lattices are asymmetric, i.e., $V(u)\neq V(-u)$, while the interactions in other lattices are symmetric. In the former case, the temperature pressure is nonvanishing, finite in the thermodynamic limit~\cite{PhysRevLett.84.2857}.

\subsubsection{Non-equilibrium heat bath method}

Firstly, we calculate the thermal conductivity $\kappa_{\rm NE}$ according to the definition Eq.~(\ref{eq:kne}) with the non-equilibrium heat bath method. The subscript `NE' indicates that the calculation is in non-equilibrium steady states.
To this end, fixed boundary conditions are applied, i.e., $u_0=u_N=0$, with $N$ being the total number of particles.
Since the lattice constant $a$ has been set as unity in the dimensionless units, the lattice length $L$ is simply equivalent to the particle number $N$ as $L=Na=N$. Two Langevin heat baths with temperatures $T=0.5$ and $1.5$ are coupled to the two ends of the lattice, respectively.
The equation of motion of the particle coupled to the heat bath is described by the following Langevin dynamics
\begin{align} \label{LE}
  \ddot u = f -\lambda \dot u + \xi,
\end{align}
where $f$ denotes the interaction force generated from other particles, $\xi$ denotes a Wiener process with zero mean and variance $2\lambda k_BT$,
and $\lambda$ denotes the relaxation coefficient of the Langevin heat bath.
Generally, the resulting heat flux approaches zero in both limits $\lambda \rightarrow 0$ and $\lambda \rightarrow \infty$~\cite{PhysRep.377.1}.
In practice, the $\lambda$ has been optimized to be $0.2$ so as to maximize the heat currents.
In order to achieve better performance, we usually put more than one particles into the heat bath in each end ~\cite{PhysRevLett.84.2381,EurophysLett.93.54002}.

\begin{figure}[ht]
\includegraphics[width=\columnwidth]{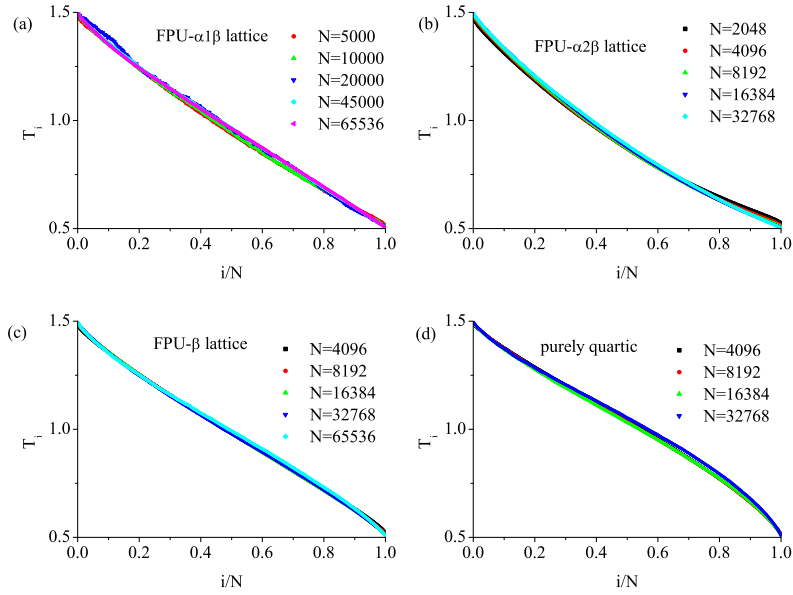}
\caption{\label{fig:Tdis1d} Temperature profiles for (a) the FPU-$\alpha1\beta$, (b) the FPU-$\alpha2\beta$ (c) the FPU-$\beta$,
                           and (d) the purely quartic lattices with various length $L$.
Only the temperature profiles in the central region are taken into account in calculating the temperature gradient $\frac{dT}{dL}$, i.e.,
the left and right 1/4 of the lattices are excluded in order to remove  boundary effects.}
\end{figure}

To avoid the effect of possible slow convergence
\cite{JPhysAMathTheor.43.145001}, the simulations have been performed long enough time until the temperature profiles are well established and
the heat currents along the lattice become constant, see Fig.\ref{fig:Tdis1d}.
The temperature gradient $\nabla T\equiv\frac{dT}{dL}$ is calculated by linear least squares
fitting of the temperature profile in the central region, aiming to greatly reduce the boundary effects.

\begin{figure}[ht]
\sidecaption[t]
\includegraphics[width=0.64\columnwidth]{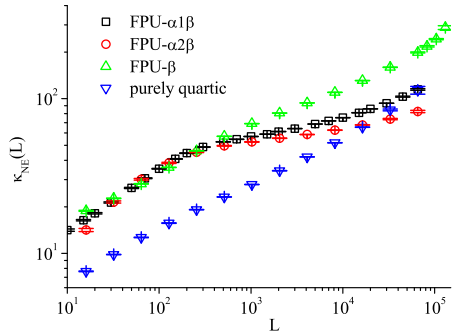}
\caption{\label{fig:Kne1d}
Thermal conductivity $\kappa_{\rm NE}$ versus lattice length $L$. The reversed tendency can be roughly seen in
the rightmost part of the figure.
}
\end{figure}

This so evaluated thermal conductivity $\kappa_{\rm NE}(L)$ are plotted for different 1D lattices in Fig.~\ref{fig:Kne1d}.
For the lattices with asymmetric interactions, a very flat length-dependence of thermal conductivity is observed with $L$ ranging from several hundreds to ten thousands sites. However, for even longer lengths, say, $L>1\times10^4$, the running slope of the thermal conductivity $\kappa_{\rm NE}(L)$ as the function of $L$ starts to grow again. By comparing the results in the two FPU-$\alpha\beta$ lattices, we see that the asymptotic tendency of curving up of $\kappa_{\rm NE}(L)$ is not affected even for the case of strong asymmetry ($k_3=2$ is the maximum value that keeps the potential single well, at the given $k_2=k_4=1$.). With the increase of asymmetry, the tendency of curving up of $\kappa_{\rm NE}(L)$ can only be postponed as shown in Fig.~\ref{fig:Kne1d}. Such a phenomenon can also be observed in the FPU-$\beta$ lattice, although the effect is much more slight. Because of this, the thermal conductivity in this lattice displays a a little bit slower divergence as $L^{1/3}$ in a certain length regime~\cite{PhysRevLett.98.184301}. This finite-size Fourier-like behavior has been repeatedly observed recently in many 1D lattices ~\cite{EurophysLett.93.54002,PhysRevE.85.060102,PhysRevE.88.052112,PhysRevE.89.032102,JStatPhys.154.204,PhysRevE.90.032134},
its physical reason is, however, still not clear.

\subsubsection{Green-Kubo scheme}

The major drawback of the non-equilibrium heat bath method is the boundary effects due to coupling with heat baths which are difficult to be sufficiently removed. In addition, the temperature difference cannot be set as small values in this method, otherwise the net heat currents can hardly be distinguished from the statistical fluctuations. Therefore, the systems prepared in the non-equilibrium heat bath method are far from ideal close-to-equilibrium states. Numerical difficulties also prevent us from simulating even longer lattices. We thus turn to calculate the equilibrium heat current autocorrelation functions with the Green-Kubo method, which provides an alternative way of determining the divergency exponent $\alpha$~\cite{GREENKUBO}. No heat bath enters into the lattice dynamics.

In a finite lattice with $N$ particles, the  heat current correlation function $c_N(\tau)$ is defined as
\begin{align}
c_N(\tau)\equiv \frac1{N}\langle J(t)J(t+\tau)\rangle_t.
\label{eq:JJ}
\end{align}
where $\langle\cdot\rangle$ denotes the ensemble average, which is equivalent to the time average for chaotic and ergodic systems considered here. Compared with $c_N(\tau)$ for finite lattice, its value in thermodynamic limit is much more meaningful, i.e.,
\begin{align}
c(\tau)\equiv\lim_{N\rightarrow\infty}c_N(\tau).
\end{align}
According to the Green-Kubo formula~\cite{GREENKUBO}, the thermal conductivity $\kappa_{\rm GK}$ is integrated as
\begin{align}
 \kappa_{\rm GK} \equiv \frac1{k_BT^2} \int_0^{\infty} c(\tau) d\tau.
\end{align}
The Boltzmann constant $k_B$ is set to unity in the dimensionless units. But $k_B$ is kept in formulas for the completeness of understanding. For anomalous heat conduction, the above integral does not converge due to the slow time decay of $c(\tau)$ in asymptotic limit. In common practice, the length-dependent thermal conductivity is calculated by introducing a
cutoff time $t_s=L/v_s$, instead of infinity, as the upper limit of the integral, i.e.,
\begin{align}
 \kappa_{\rm GK}(L) \equiv \frac1{k_BT^2} \int_0^{L/v_s} c(\tau) d\tau,
\label{eq:GKL}
\end{align}
where the constant $v_s$ is the speed of sound and the subscript `GK' denotes that the calculation is based on the Green-Kubo formula. 
Since we are only interested in the divergency exponent $\alpha$ of $\kappa_{\rm GK}(L)$, its exact value is not relevant to any conclusion we made.

However, in numerical calculations, only lattices with finite $N$ can possibly be simulated. The $c_N(\tau)$ generally depends on the lattice length $N$, and the finite-size effects must be taken into consideration very carefully. We next present the simulation with a very long lattice length of $N=20000$ followed by the discussion of finite-size effects.

The simulations are carried out in lattices with periodic boundary conditions, which is known to
provide the best convergence to thermodynamic limit properties. Microcanonical simulations are performed with zero total
momentum and identical energy density $\epsilon$ which corresponds to the same temperature $T=1$ for all lattices.
The energy density $\epsilon$ equals to $0.864, 0.846, 0.867$ and $0.75$,
for the 1D FPU-{$\alpha1\beta$} lattice, the FPU-{$\alpha2\beta$} lattice, the FPU-{$\beta$} lattice, and the purely quartic lattice, respectively.

\begin{figure}[ht]
\includegraphics[width=\columnwidth]{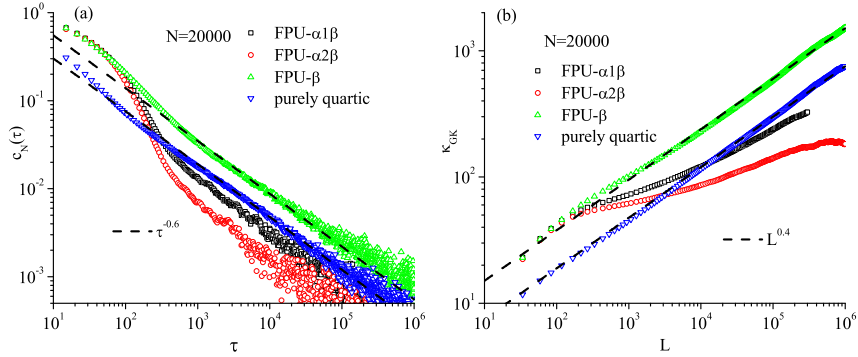}
\caption{\label{fig:gk1d} (a) The heat current correlation $c_N(\tau)$ versus time lag $\tau$.
               Lines with slope $-3/5$ are drawn for reference.
               Data for the FPU-$\beta$ and the purely quartic lattices fit them quite well in the long $\tau$ regime.
               (b) $\kappa_{\rm GK}(L)$. Lines with slope $2/5$ are drawn for reference.
               For the FPU-$\beta$ lattice, the slope $2/5$ fits very well in the regime of $L$ from $10^3$ to $10^6$.
               It fits for the purely quartic lattice in even wider regime from $10^2$ to $10^6$.
               }
\end{figure}

The time $\tau$ dependence of $c_N(\tau)$ are shown in Fig.~\ref{fig:gk1d}(a). In a very large window of $\tau$, the $c_N(\tau)$ for the FPU-$\beta$ lattice and the purely quartic lattice follows a power-law decay of $c_N(\tau)\propto\tau^{\gamma}$ with $\gamma=-3/5$ very well. It should be pointed out that the decay exponent $\gamma$ is related to the divergency exponent $\alpha$ as $\gamma=\alpha-1$ resulted from Eq. (\ref{eq:GKL}).
While for the FPU-$\alpha\beta$ lattices, the $c_N(\tau)$ decays very fast before it approaches an asymptotic power-law decay behavior.

The corresponding length dependence of the thermal
conductivity $\kappa_{\rm GK}(L)$ from Eq. (\ref{eq:GKL}) is plotted in Fig.~\ref{fig:gk1d}(b).
For the FPU-$\beta$ lattice, the best fit of the data from $L=10^4$ upward gives rise to a divergency exponent $\alpha=0.42$, which strongly prefers the theoretical prediction of $\alpha=2/5$ to $\alpha=1/3$. As for the purely quartic lattice, the best fit of the data for $L$ from $10^2$ upward, covering four orders of magnitude, yields $\alpha=0.41$, which is even closer to the prediction of $\alpha=2/5$.

\begin{figure}[ht]
\sidecaption[t]
\includegraphics[width=0.64\columnwidth]{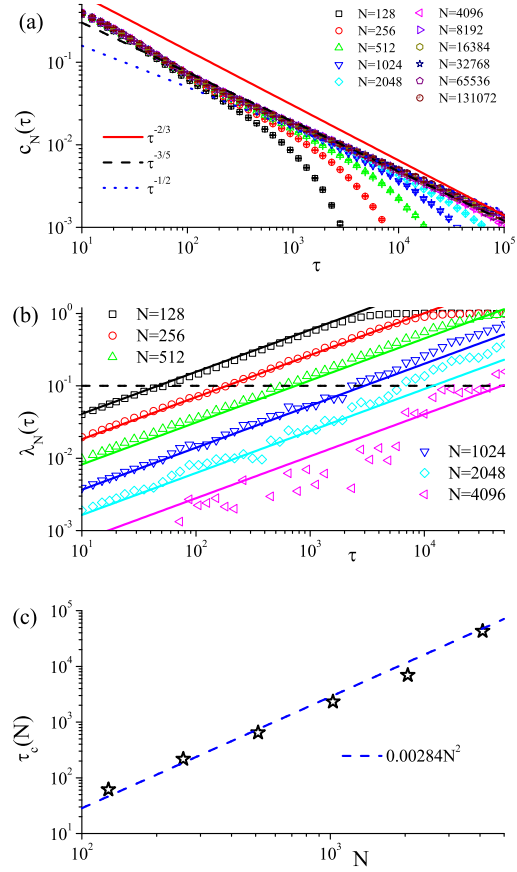}
\caption{\label{fig:correlationloss}
(a) The heat current correlation $c_N(\tau)$ for the purely quartic lattice for various lattice length $N$.
Lines with different slopes $-2/3$, $-3/5$ and $-1/2$, respectively, are drawn for reference.
  (b) $\lambda_N(\tau)$ as a function of $\tau$ for various lattice length $N$.
 Oblique solid lines from the top down stand for the fittings of $\lambda_N(\tau)$, $3N^{-1.16}\tau^{0.58}$, for $N$ ranging from 128 to 4096.
 The horizontal dashed line refers to $\lambda_N(\tau)=0.1$.
 $\lambda_N(\tau)$ cross this line at the cut off time lag $\tau_c(N)$.
 (c) The cutoff time lag $\tau_c(N)$ as a function of the lattice length $N$. The blue dashed line stands for the expectation in Eq.~(\ref{tauc}), $2.84\times 10^{-3}N^2$.}
\end{figure}

{\it Heat current correlation loss}.
To quantitatively evaluate the finite-size effects,
we plot the heat current correlation $c_N(\tau)$ for the 1D purely quartic lattice with sizes from $N=128$ to $N=131072$ in Fig.~\ref{fig:correlationloss}(a).
We define a heat-current correlation loss (CCL) $\Delta c_N(\tau)$ induced by the finite-size effects as
\begin{align}
\Delta c_N(\tau)\equiv c(\tau)-c_N(\tau).
\end{align}
The relative loss $\lambda_N(\tau)$ is defined as
\begin{align}
\lambda_N(\tau)\equiv \frac{\Delta c_N(\tau)}{c(\tau)}=1- \frac{c_N(\tau)}{c(\tau)}.
\end{align}
By setting a certain critical value $\eta$ for the relative loss $\lambda_N(\tau)$, a characteristic time lag, the cutoff time lag $\tau_c(N)$ can be obtained for each length $N$ by solving,
\begin{align}\label{eq:cutoff-time}
\lambda_N(\tau_c(N))=\eta.
\end{align}

Since the asymptotic $c(\tau)$ can never be actually calculated, we thus need to approximately replace $c(\tau)$ with $c_N(\tau)$ for a finite long enough lattice instead.
Under any existing criterion~\cite{PhysRevE.89.022111}, the length of $N=131072$ is long enough for correlation times $\tau\leq5\times10^4$.
Therefore, the asymptotic $c(\tau)$ refers to $c_{131072}(\tau)$ in the descriptions of our numerical simulations hereafter.

In Fig.~\ref{fig:correlationloss}(b), the relative loss $\lambda_N(\tau)$ for the purely quartic lattice for various length $N$ is plotted .
The $\lambda_N(\tau)$ is larger in shorter lattices as one should expect. Interesting enough, all the data of $\lambda_N(\tau)$ as the correlation time $\tau$ for various $N$ fit the following universal relation quite well:
\begin{align}\label{fitlambda}
\lambda_N(\tau)\approx3N^{-1.16}\tau^{0.58}.
\end{align}
This relation implies that the cutoff time lag $\tau_c(N)$ should follow a square-law dependence on the lattice length $N$:
\begin{align} \label{tauceta}
\tau_c(N)\approx (\frac{\eta}3)^{\frac1{0.58}}N^2.
\end{align}
For the critical value of $\eta=0.1$, namely, the $c_N(\tau)$ decreases to $90\%$ of the value of $c(\tau)$ at $\tau=\tau_c(N)$.
 The cutoff time lag $\tau_c(N)$ as the function of length $N$ can be obtained from Eq.~(\ref{eq:cutoff-time}) as
\begin{align} \label{tauc}
\tau_c(N)\approx 2.84\times 10^{-3}N^2.
\end{align}
This is shown in Fig.~\ref{fig:correlationloss}(c) where good agreement with numerical data can be observed.

It is reasonable to expect that Eq.~(\ref{fitlambda}) should also remain valid in larger $N$ regime.
We are thus able to estimate the value of relative loss $\lambda_N(\tau)$ in Fig.~\ref{fig:gk1d}(a), which is no more than $10\%$.
Given the fact that $c_N(\tau)\propto\tau^{\gamma}$ was fitted over four orders of magnitude of $\tau$,
the underestimate of $\delta$ induced therefrom must not be higher than $|\log_{10}0.9|/4\approx0.01$.
It is noticed that such an error is much smaller than the difference between the three theoretical expectations $\gamma=-2/3$, $\gamma=-3/5$ and $\gamma=-1/2$.
The conclusion that $c(\tau)$ is best fitted as $c(\tau)\propto\tau^{-3/5}$ should not be affected by this finite-size effect.
We also expect that the cases in 2D~\cite{PhysRevE.86.040101} and 3D~\cite{PhysRevLett.105.160601}
purely quartic lattices are also similar.

In summary, we have numerically calculated the length-dependent thermal conductivities $\kappa$ in a few typical 1D lattices by using both non-equilibrium heat bath and equilibrium Green-Kubo methods. Consistent results are obtained for thermal conductivity divergency exponent $\alpha$.
For the FPU-$\beta$ and the purely quartic lattices, the thermal conductivities $\kappa$ follows a power-law length-dependence of $\kappa\propto L^{0.4}$ very well, for a wide regime of $L$. While for the asymmetric FPU-$\alpha\beta$ lattices, large finite-size effects are observed. As a result, $\kappa$ increases with lattice length very slowly in a wide range of $L$.
Our numerical simulations indicate that $\kappa$ regains its increase in yet longer length $L$~\cite{EurophysLett.93.54002,PhysRevE.88.052112}.
This is also consistent with some recent studies~\cite{PhysRevE.89.032102,JStatPhys.154.204}.

The studies of the heat current correlation loss in the purely quartic lattice indicate that for a not-very-large $N$,
$c_N(\tau)$ is close enough to the asymptotic $c(\tau)$ within a very long correlation time $\tau$ window.
Therefore, we are able to extract  $\kappa_{\rm GK}(L)$ from Eq. (\ref{eq:GKL}) with an effective very long $L$
by performing simulations in a lattice with relatively much smaller size $N$, i.e. $L=v_s \tau_c>>N$. The research area of the investigation of heat transport applicable for the Green-Kubo method is thus greatly broadened~\cite{PhysRevE.91.012110}.

\subsection{Logarithmic divergent thermal conductivity in 2D momentum-conserving nonlinear lattices}\label{sec:2d}

For 2D and 3D momentum-conserving systems with higher dimensionality, consistent predictions are achieved from different theoretical approaches. The linear response approaches based on the renormalization group~\cite{PhysRevLett.89.200601} and mode-coupling
theory~\cite{PhysRep.377.1,PhysRevA.4.2055,JStatPhys.15.7,JStatPhys.15.23} both predict that the heat current
autocorrelation function $c(\tau)$ decays with the correlation time $\tau$ as
$c(\tau)\propto\tau^{\gamma}$, where $\gamma={-1}$ and ${-3/2}$ for 2D and 3D, respectively.
These predictions indicate that the resulted thermal conductivity is logarithmically divergent as $\kappa\propto \ln{L}$ in 2D systems
and a finite value in 3D systems.

In 2D momentum-conserving systems, a logarithmic divergence of thermal conductivity of $\kappa\propto \ln{L}$ is reported by numerical simulations in the FPU-$\beta$ lattice with rectangle~\cite{JStatPhys.100.1147,PhysRevE.74.062101} and disk~\cite{PhysRevE.82.030101} geometries where vector displacements are considered.
However, a power-law divergent thermal conductivity of $\kappa\propto L^{\alpha}$ is also observed in the 2D FPU-$\beta$ lattices with scalar displacements~\cite{2002cond.mat4247G}.

\begin{figure}[ht]
\sidecaption[t]
\includegraphics[width=0.64\columnwidth]{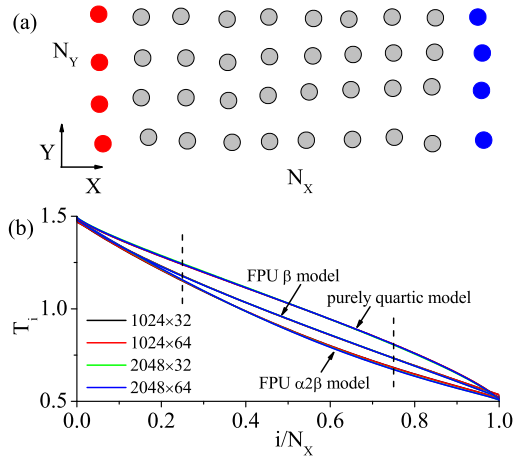}
\caption{\label{fig:tdis2d} (color online).
(a) Scheme of a 2D square lattice. Heat current along the $X$ axis is calculated.
(b) Temperature profiles of different lattices.
Curve groups from down top correspond to the FPU-$\alpha2\beta$, FPU-$\beta$, and purely quartic lattices, respectively.
Only the data in the central region between the
two vertical dashed lines are taken into account in calculating the temperature gradient $\nabla T$.
}
\end{figure}

In this subsection we systematically study the heat conduction in a few 2D square lattices with a scalar
displacement field $u_{i,j}$, where the schematic 2D setup is plotted in Fig.~(\ref{fig:tdis2d})(a).
The scalar 2D Hamiltonian reads
\begin{eqnarray}\label{eq:Ham2d}
H=\sum_{i=1}^{N_X}\sum_{j=1}^{N_Y} [ \frac{p^2_{i,j}}2 + V(u_{i,j}-u_{i-1,j})+V(u_{i,j}-u_{i,j-1}) ],
\end{eqnarray}
where $N_X$ and $N_Y$ denotes the number of layers in $X$ and $Y$ directions. The inter-particle potential takes the FPU form of $V(u)=\frac{1}{2}k_2u^2+\frac{1}{3}k_3u^3+\frac{1}{4}k_4u^4$.
Dimensionless units is applied and all the particle masses has been set to unity.
In order to justify the logarithmic divergence and also study the influence of inter-particle coupling,
we choose three types of lattices, i.e., the FPU-$\alpha2\beta$ lattice: $k_2=k_4=1$, $k_3=2$;
the FPU-$\beta$ lattice: $k_2=k_4=1$, $k_3=0$; and the purely quartic lattice: $k_2=k_3=0$, $k_4=1$.

The interaction forces between a particle labeled $(i,j)$ and its nearest right and up neighbors
are $f^X_{i,j}=-{d V(u_{i,j}-u_{i-1,j})}/{d u_{i,j}}$ and
$f^Y_{i,j}=-{d V(u_{i,j}-u_{i,j-1})}/{d u_{i,j}}$. The local heat currents in the two directions are defined as
$j^X_{i,j}=\frac12 (\dot u_{i,j}+\dot u_{i+1,j})f^X_{i+1,j}$ and
$j^Y_{i,j}=\frac12 (\dot u_{i,j}+\dot u_{i,j+1})f^Y_{i,j+1}$, respectively.

\subsubsection{Non-equilibrium heat bath method}

\begin{figure}[ht]
\includegraphics[width=\columnwidth]{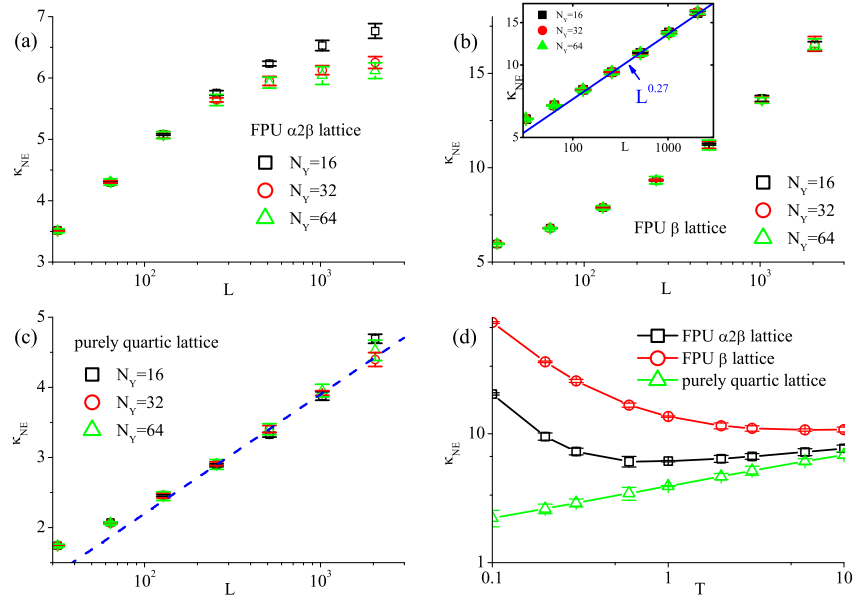}
\caption{\label{fig:kne2d}
Thermal conductivity
$\kappa_{\rm NE}$ in 2D (a) the FPU-$\alpha\beta$, (b) the FPU-$\beta$ and
(c) the purely quatic lattices versus lattice length $L$ for various $N_Y$.
The dashed line that indicate logarithmic growth is drown for reference.
Inset of (b): data for the FPU-$\beta$ lattice in double logarithmic scale.
Solid line corresponds to the power-law divergence $L^{0.27}$.
(d)$\kappa_{\rm NE}$ versus temperature $T$ in various lattices.
}
\end{figure}

We first calculate the thermal conductivity $\kappa_{\rm NE}$ in
non-equilibrium stationary states. The fixed boundary conditions are applied in the $X$-direction,
while periodic boundary conditions are applied in the $Y$-direction.
The left- and right- most columns are coupled to Langevin heat baths with temperatures $T_L=1.5$ and $T_R=0.5$, respectively, see Fig.~(\ref{fig:tdis2d})(a).
Heat currents through the $X$-direction along with the direction of temperature gradient are measured.
For each lattice with the largest size ($2048\times64$),
the average heat current is performed over time period of $2\sim4\times 10^7$ in dimensionless units after long enough transient time.
The temperature of each column is defined as the average temperature of all the particles in that column, i.e.,
\begin{equation}
T(i)\equiv \frac1{N_Y} \sum_{j=1}^{N_Y} T(i,j) = \frac1{N_Y} \sum_{j=1}^{N_Y} \langle\dot u^2_{i,j}\rangle,  \nonumber
\end{equation}
The temperature profiles of different lattices for various $N_X\times N_Y$ are plotted in Fig.~\ref{fig:tdis2d}(b).
Those profiles with different sizes for the same lattice are all overlapped with each other,
which indicates that the temperature gradient $\nabla T$ along the $X$-direction can be well established.
It is also confirmed that the temperature profiles and the heat currents along the lattices approach constant values which are independent of the overall time used here. The thermal conductivity $\kappa$ for 2D systems is defined as:
\begin{equation}
\kappa_{\rm NE} = -\frac {\langle J\rangle}{N_Y \nabla T},  \nonumber
\end{equation}
where $J$ stands for the total heat current, and the temperature gradient $\nabla T$ is along the X direction.
Since the lattice constant $a$ is set to unity, the lattice length $L$ is simply equivalent to the number of layers $N_X$ in $X$-direction.
As shown in Fig.~\ref{fig:tdis2d}(b), the shapes of temperature profiles are obviously nonlinear. Such a nonlinearity is caused by boundary effects rather than the intrinsic temperature dependence of the thermal conductivity, as can be concluded by examining the temperature dependence of $\kappa_{\rm NE}$ for different lattices in Fig.~\ref{fig:kne2d}(d). To reduce this boundary effect, the temperature gradient $\nabla T$ is calculated by a linear least-squares fitting of the temperature profiles in the central region where the left- and right- most 1/4 of the lattices are excluded.

In Fig.~\ref{fig:kne2d}(a) to (c), the thermal conductivities $\kappa_{\rm NE}$ versus $L$ with different widths of $N_Y$ are plotted in linear-log scales. In the large lattice size region, the narrow lattices with smaller $N_Y$ posses higher values of thermal conductivities $\kappa_{\rm NE}$. This is not a surprise since the narrow 2D lattices are much closer to a 1D lattice where high thermal conductivities are expected. The length dependence of $\kappa_{\rm NE}$ for the 2D FPU-$\alpha2\beta$ lattice (Fig.~\ref{fig:kne2d}(a)) with $N_Y=64$ becomes flat, indicating that $\kappa_{\rm NE}$ increases much more slowly than a logarithmic growing. In contrast, the thermal conductivities $\kappa_{\rm NE}$ for the 2D FPU-$\beta$ lattice diverges with length $L$ more rapidly than a logarithmic divergence, as can seen from Fig.~\ref{fig:kne2d}(b) and its inset.
This is actually a power-law divergence of $\kappa\propto L^{\alpha}$ and the divergency exponent $\alpha$ can be estimated from the best fit of the last four points as $\alpha=0.27\pm0.02$. However, for 2D purely quartic lattices, the $\kappa_{\rm NE}$ displays a logarithmic growth as $\kappa\propto \ln{L}$ over at least one order of magnitude in length scale, as in Fig.~\ref{fig:kne2d}(c).

\subsubsection{Green-Kubo method}

Similar to the situation in 1D lattices~\cite{EurophysLett.93.54002},
finite-size and finite-temperature-gradient effects from the non-equilibrium heat bath method are considerable and not easy to be removed.
We shall turn to the equilibrium Green-Kubo method~\cite{GREENKUBO} to seek higher accuracy of numerical results.

In the Green-Kubo simulation, periodic boundary conditions are applied in
both the $X$- and $Y$- directions.
The autocorrelation function of total heat current $c_{N_X,N_Y}^X(\tau)$ in the $X$ direction is defined as
\begin{equation}
c_{N_X,N_Y}^X(\tau)\equiv \frac1{N_XN_Y}\langle J^X(t)J^X(t+\tau)\rangle_t,
\label{eq:JJ2d}
\end{equation}
where $J^X(t)\equiv\sum_{i,j} j^X_{i,j}(t)$ is the instantaneous
total heat current in the $X$-direction. 
For simplicity, the subscripts $N_X$ and $N_Y$ of $c_{N_X,N_Y}^X(\tau)$ are omitted as $c^X(\tau)$ hereafter except in case of necessity.
The length-dependent thermal conductivity $\kappa_{\rm GK}(L)$ from the Green-Kubo method can be defined as
\begin{equation}
 \kappa_{\rm GK}(L) \equiv \frac1{k_BT^2} \lim_{N_X\rightarrow\infty}\lim_{N_Y\rightarrow\infty} \int_0^{L/v_s} c^X(\tau) d\tau,
\label{eq:GK2d}
\end{equation}
where $v_s$ is again the speed of sound.
Microcanonical simulations are performed with zero total momentum~\cite{PhysRep.377.1} and specified energy density
$\epsilon$ which corresponds to the same temperature $T=1$ for different lattices. The energy density
$\epsilon$ equals 0.887, 0.892 and 0.75 for the 2D FPU-$\alpha2\beta$, FPU-$\beta$, and
purely quartic lattices, respectively.
A number of independent runs (64 for $1024\times1024$ and fewer for smaller lattices) are carried out.
Simulations of the largest lattices ($1024\times1024$) are performed for about total time of $10^7$ in the dimensionless units.

\begin{figure}[ht]
\includegraphics[width=0.5\columnwidth]{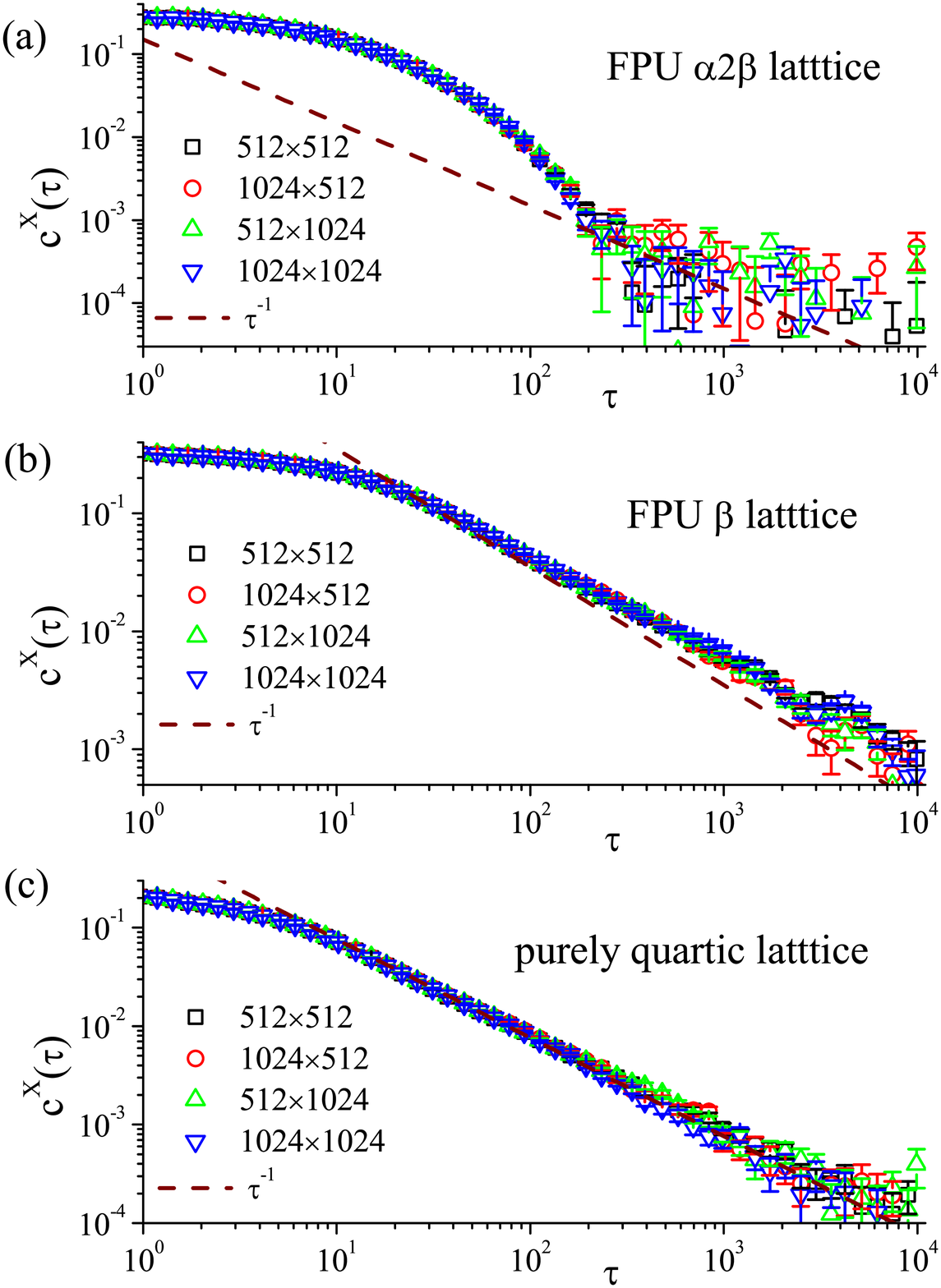}
\includegraphics[width=0.5\columnwidth]{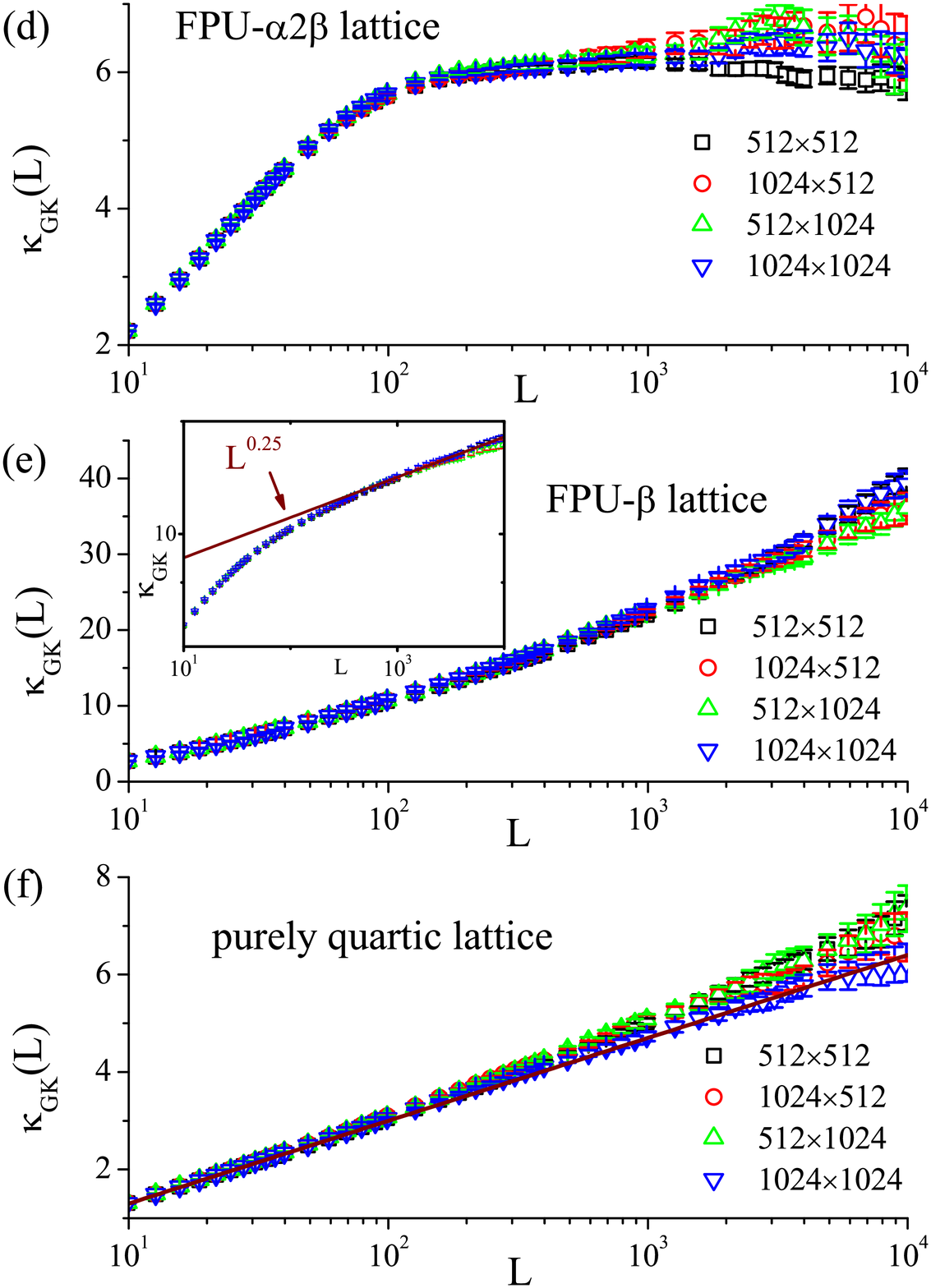}
\caption{\label{fig:gk2d}
(a) to (c), $c^X(\tau)$ for in $N_X\times N_Y$ lattices.
Lines correspond to $\tau^{-1}$ are drawn for reference.
(a) The FPU-$\alpha2\beta$ lattice. $c^X(\tau)$ decays much faster than $\tau^{-1}$ in short $\tau$ regime.
The decay tend to slow down for $\tau>300$.
(b) The FPU-$\beta$ lattice. $c^X(\tau)$ decays obviously slower than $\tau^{-1}$ in a quite wide regime of $\tau$ .
(c) The purely quartic lattice. $c^X(\tau)$ follows $\tau^{-1}$ very well in nearly three orders of magnitude of $\tau$.
(d) to (f), $\kappa_{\rm GK}(L)$ in the $X$-direction in $N_X\times N_Y$ lattices.
(d) The FPU-$\alpha2\beta$ lattice. A flat $\kappa_{\rm GK}$ is again observed for $L<2000$.
    Thereafter $\kappa$ tend to rise up.
    It is easy to understand that slow down of the decay of $c(\tau)$ cannot instantly induce a visible
    rise up of $\kappa_{\rm GK}$, since $c(\tau)$ has already decayed to a too low value.
(e) The FPU-$\beta$ lattice. In a wide regime of $\tau$, $c_{\tau}$ decays obviously slower than $\tau^{-1}$.
    Inset: data plotted in double logarithmic scale. Solid line corresponds to $L^{0.25}$.
(f) The purely quartic lattice.
    $\kappa_{\rm GK}$ for $1024\times1024$ follows the straight line very well in nearly three orders of magnitude of $\tau$,
    This strongly supports a logarithmically divergent thermal conductivity.
    The slight rise for smaller lattices is due to the finite-size effect.
}
\end{figure}

The decays of $c^X(\tau)$ with the correlation time $\tau$ for different 2D lattices are plotted in Fig.~\ref{fig:gk2d} (a) to (c).
To eliminate the finite-size effects, we have performed simulations by varying $N_X$ and $N_Y$ and only consider
the asymptotic behavior which is the part of curves overlapping with each other.
Within the range of standard error, the satisfactory overlap with each other is clearly observed.
In the specific cases with $N_X=N_Y$, the average of the autocorrelation function of $[c^X(\tau)+c^Y(\tau)]/2$ is plotted instead.
This is equivalent to double the simulation time steps to achieve higher accuracy without actually performing any more computation.
And due to the symmetry of the square lattice, it is obvious to find that $c_{512,1024}^X(\tau)=c_{1024,512}^Y(\tau)$, where the simulations for the two lattices can be carried out in the same run.

It is observed that in Fig.~\ref{fig:gk2d}(a), the $c^X(\tau)$ in the 2D FPU-$\alpha2\beta$ lattices decays much faster than the theoretical prediction of $c^X(\tau)\propto\tau^{-1}$ in a wide regime of correlation time $(\tau)$. As a result, the integrated $\kappa_{\rm GK}$ displays a saturation behavior with the length for large $L$ in Fig.~\ref{fig:gk2d}(d). The rapid decay of $c^X(\tau)$ tends to slow down for yet longer time $\tau$. However, the asymptotic behavior cannot be numerically confirmed due to large fluctuations.
In Fig.~\ref{fig:gk2d}(b), the $c^X(\tau)$ in the 2D FPU-$\beta$ lattice decays evidently more slowly than $\tau^{-1}$, which gives rise to a power-law divergence of thermal conductivity of $\kappa_{\rm GK}\propto L^{\alpha}$ in Fig.~\ref{fig:gk2d}(e).
The best fit of this divergency exponent $\alpha$ in the regime of $L>10^3$ is obtained as $\alpha=0.25\pm0.01$.
For the 2D purely quartic lattice as seen in Fig.~\ref{fig:gk2d}(c), the $c^X(\tau)$ decays as the predicted behavior of $c^X(\tau)\propto\tau^{-1}$ for nearly three orders of magnitudes of correlation time $\tau$.
This finding strongly supports a logarithmic diverging thermal conductivity of $\kappa\propto \ln{L}$, which can be clearly observed in Fig.~\ref{fig:gk2d}(f). In all cases, the tendency of $\kappa_{\rm GK}$ from Green-Kubo method (shown in the right column of Fig.~\ref{fig:gk2d})
 is in good agreement with that of $\kappa_{\rm NE}$ from non-equilibrium heat bath method (shown in Fig.~\ref{fig:kne2d}).

In summary, we have extensively studied heat conduction in three 2D nonlinear lattices with both non-equilibrium heat bath method and Green-Kubo method. The roles of harmonic and asymmetric terms of the inter-particle coupling are clearly observed by
comparing the results for the purely quartic lattice and the other two latices.
In the 2D purely quartic lattice, the heat current autocorrelation function $c(\tau)$
is found to decay as $\tau^{-1}$ in three orders of magnitude from $10^1$ to $10^4$.
This strongly supports a logarithmically divergent thermal conductivity of $\kappa\propto \ln{L}$ consistent with the theoretical predictions.
For the 2D FPU-$\beta$ lattice, our non-equilibrium and equilibrium calculations suggest a power-law divergence with a divergency exponent
$\alpha=0.27\pm0.02$ and $0.25\pm0.01$, respectively. A very significant finite-size effect which results a flat length dependence of $\kappa(L)$ is observed in the 2D FPU-$\alpha\beta$ lattice
with asymmetric potential.
Most existing numerical studies on 2D lattices with asymmetric interaction terms
suggest a logarithmically divergent behavior as $\kappa\propto \ln{L}$, e.g., the Fermi-Pasta-Ulam (FPU)-$\beta$ lattice with
rectangle~\cite{JStatPhys.100.1147,PhysRevE.74.062101}
and disk~\cite{PhysRevE.82.030101} geometries.
It might be due to that the effect of the harmonic term is largely offset by that of the asymmetric term,
thus yielding a logarithmic-like divergence of thermal conductivity.

Similar to the findings of 1D lattices where $\kappa$ tends to diverge with length in the same way in the
thermodynamic limit for all kinds of lattices~\cite{EurophysLett.93.54002},
it should also be expected that $\kappa$ will diverge as $\log L$ in long enough 2D FPU-$\alpha\beta$ and FPU-$\beta$ lattices as already observed for 2D purely quartic lattice. However, in order to see such an asymptotic divergence,
2D lattices with much larger sizes have to be simulated which is beyond the scope of our current studies.

We should emphasize that such numerical studies are not only of theoretical importance.
Progresses in nano-technology have made it possible to experimentally measure the size dependence of thermal conductivities
in some 1D~\cite{PhysRevLett.101.075903} and 2D~\cite{NatMater.9.555,2010arXiv1012.2937X,NanoLett.11.113,NanoLett.12.3238} nano-scale materials.

\subsection{Normal heat conduction in a 3D momentum-conserving nonlinear lattice}\label{sec:3d}

For 3D momentum-conserving systems, all the above-mentioned theories predict that the heat current
autocorrelation function decays with correlation time $\tau$ as
$\tau^{\gamma}$ with ${\gamma=-3/2}$, which gives rise to a normal heat conduction.
However, numerical simulations are not so conclusive
\cite{JPhysSocJpn.75.103001,JPhysSocJpn.77.054006}.
It is only reported in 2008, by non-equilibrium simulations,
that the running divergency exponent $\alpha_L \equiv d
\ln \kappa/d \ln L$ of the 3D FPU-$\beta$ lattice shows
a power law decay in $L$, thus vanishes in the
thermodynamic limit~\cite{PhysRevLett.104.040601}.
Normal heat conduction in 3D systems is therefore verified. However, according to the Green-Kubo formula, any power-law decay of $c(\tau)\propto t^{\gamma}$ with ${\gamma}<-1$ will yield a finite value of thermal conductivity signaturing a normal heat conduction behavior~\cite{GREENKUBO}. In order to confirm the theoretical prediction of the specific value of ${\gamma}=-3/2$, the heat current autocorrelation function $c(\tau)$ must be directly calculated by using the equilibrium Green-Kubo method.

We investigate the decay of the heat current
autocorrelation function in a 3D cubic lattice with a scalar
displacement field $u_{i,j,k}$. The 3D Hamiltonian reads
\begin{eqnarray}\label{eq:Ham3d}
H=\sum_{i=1}^{N_X}\sum_{j=1}^{N_Y}\sum_{k=1}^{N_Z}&&\left[\frac{p^2_{i,j,k}}2 + V(u_{i,j,k}-u_{i-1,j,k})\right.\nonumber\\
&&\left.+V(u_{i,j,k}-u_{i,j-1,k})+V(u_{i,j,k}-u_{i,j,k-1})\right],
\end{eqnarray}
where $V(u)=\frac14u^4$ takes the purely quartic form. We choose this purely quartic potential
due to its simplicity and high nonlinearity, where close-to-asymptotic behaviors can be achieved in shorter time and space scales.
This model can also be regarded as the high temperature limit of the FPU-$\beta$ model.
Periodic boundary conditions, which
provide the best convergence to thermodynamic limits, are applied in
all three directions, i.e., $u_{N_{X},j,k}= u_{0,j,k}$, $u_{i,{N_Y},k}= u_{i,0,k}$ and $u_{i,j,{N_Z}}=u_{i,j,0}$.
The interactions between a particle $(i,j,k)$ and its nearest neighbors
are: $f^X_{i,j,k}=-{d V(u_{i,j,k}-u_{i-1,j,k})}/{d u_{i,j,k}}$,
$f^Y_{i,j,k}=-{d V(u_{i,j,k}-u_{i,j-1,k})}/{d u_{i,j,k}}$ and
$f^Z_{i,j,k}= \\ -{d V(u_{i,j,k}-u_{i,j,k-1})}/{d u_{i,j,k}}$. The local
heat current in three directions are defined as
$j^X_{i,j,k}=\frac12 (\dot u_{i,j,k}+\dot u_{i+1,j,k})f^X_{i+1,j,k}$,
$j^Y_{i,j,k}=\frac12 (\dot u_{i,j,k}+\dot u_{i,j+1,k})f^Y_{i,j+1,k}$, and
$j^Z_{i,j,k}=\frac12 (\dot u_{i,j,k}+\dot u_{i,j,k+1})f^Z_{i,j,k+1}$, respectively.
For convenience and simplicity, $N_X=N_Y=W$ is always chosen and the focus is on the heat
conduction in the $Z$ direction with different cross section area of $W^2$ and length of $N_Z$.

The heat current autocorrelation function $c^Z(\tau)$ in the $Z$
direction for a given $W$ is defined as \cite{GREENKUBO}
\begin{equation}\label{eq:JJ3d}
c^Z(\tau)\equiv\lim_{N_Z\rightarrow\infty}\frac1{W^2N_Z}\langle J^Z(t)J^Z(t+\tau)\rangle_t,
\end{equation}
where $J^Z(t)\equiv\sum_{i,j,k} j^Z_{i,j,k}(t)$ is the instantaneous
total heat current in $Z$ direction.
The length-dependent thermal conductivity $\kappa^Z_{\rm GK}(L)$ is defined as
\begin{equation} \label{GK3d}
 \kappa^Z_{\rm GK}(L) \equiv  \frac1{k_BT^2} \int_0^{L/v_s} c^Z(\tau) d\tau,
\end{equation}
where the constant $v_s$ is the speed
of sound~\cite{PhysRep.377.1,AdvPhys.57.457}.
$v_s$ is of order 1 for the present lattice. 
Microcanonical simulations are performed with zero
total momentum\cite{PhysRep.377.1,EurophysLett.43.271,PhysRevE.68.067102} and fixed energy density $\epsilon=0.75$ which
corresponds to the temperature $T=1$.
Due to the presence of statistical fluctuations, the simulation must be carried out long enough, otherwise the real decay exponents of the autocorrelation function cannot be determined with good enough accuracy.
We perform the calculations by 64-thread parallel computing.
The simulation of the largest lattice ($W=64$ and $N_Z=128$) is performed for the total time of $5\times 10^6$ in dimensionless units.

\begin{figure}[ht]
\sidecaption[t]
\includegraphics[width=0.64\columnwidth]{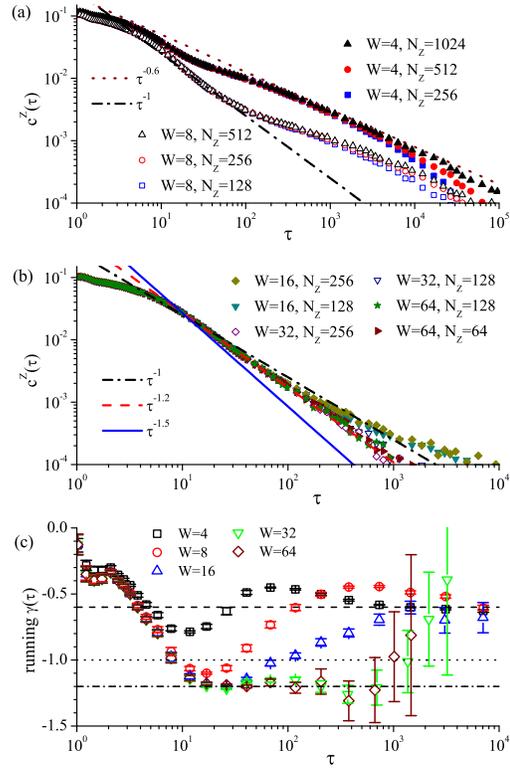}
\caption{\label{fig:3dkubojjrunningb} (color online).
(a) $c^Z(\tau)$ versus $\tau$ for various $N_Z$ and $W=4$ and $8$. Lines with slopes -0.6 and -1 are plotted for reference.
(b) $c^Z(\tau)$ versus $\tau$ for various $N_Z$ and $W=16$, $32$, and $64$. Lines with slopes -1, -1.2, and -1.5 are plotted for reference.
The larger the $W$, the longer the curve follows the power law decay $\tau^{-1.2}$, which suggests that $c(\tau)$
follows $\tau^{-1.2}$ as $W$ is large enough.
(c) Dependence of the
running exponent $\gamma(\tau) \equiv d \ln c^Z(\tau)/d \ln \tau$ on
the time lag $\tau$. Lines for $\gamma={-0.6}$, ${-1}$, and ${-1.2}$ are depicted for reference.
One can see that for $W\geq16$, the bottoms of the
curves stop decreasing and tend to saturate at a $W$-independent
value $-1.2$. The curves stay here for a longer time as $W$
increases.}
\end{figure}

The decay of the autocorrelation function $c^Z(\tau)$ with the correlation time $\tau$ is plotted in Fig.~\ref{fig:3dkubojjrunningb}(a) and (b).
For a given width $W$, we perform simulations by varying the lattice length $N_Z$ and consider only
the asymptotic behavior shown in the part of curves overlapping with
each other to avoid finite-size effects.
In the short-time region, typically, for $\tau<10^1$, all the curves of $c^Z(\tau)$ are relatively flat.
This corresponds to the ballistic transport regime when $\tau$ is shorter than or
comparable to the phonon mean lifetime. Except in this ballistic
regime, the $c^Z(\tau)$ for $W=4$ always decays more slowly than
$\tau^{-1}$, implying that the cross-section is too small to display a genuine 3D behavior.
The $c^Z(\tau)$ for $W=8$ decays faster than
$\tau^{-1}$ in the regime $\tau \in (10,30)$, showing a weak 3D
behavior. In longer time, the decay exponent $\gamma$ becomes less negative and
reverts to the exponent similar to that for the case $W=4$. For
$W=16$, the curves show a 3D behavior longer and finally again
revert to the 1D-like behavior. For width $W=32$ and 64 where the
similar reversal is expected, we fail to observe the 3D behavior in the asymptotic time limit due to the presence
of large statistical fluctuations. This picture indicates a
crossover from 3D to 1D behavior appearing at a $W$-dependent threshold for the correlation time as
$\tau_c(W)$. Below this critical time $\tau_c$, the system displays a 3D behavior, while a 1D behavior is recovered above $\tau_c$. In a macroscopic system, where the width and the length are comparable, only 3D behaviors can
be observed. This might be a consequence of the universality of
Fourier's law in the macroscopic world in nature.

Furthermore, one can see that, for $W\geq8$, the larger
the width, the longer the $c^Z(\tau)$ shows a power-law decay of
$\tau^{-1.2}$. This suggests that the asymptotic behavior of the
autocorrelation function should be $c^Z(\tau)\propto\tau^{-1.2}$ when $W$ is
large enough. It should be emphasized that the decay exponent is
different from the traditional theoretical prediction of
${\gamma}=-3/2$. Interestingly, the numerically observed ${\gamma}=-1.2$ agrees with the
formula $\gamma=-2d/(2+d)$ (for $d=3$) based on the hydrodynamic
equations for a normal fluid with an added thermal noise~\cite{PhysRevLett.89.200601}.
However, in that paper the authors limit the validity of this
formula to $d\leq2$. Our result ${\gamma}=-1.2$ is also compatible
with the value ${\gamma}=-0.98\pm0.25$ for the 3D FPU model reported
in Ref.~\cite{JPhysSocJpn.75.103001}.

In order to illustrate the 3D-1D crossover more clearly, we plot the $\tau$-dependence of the
running decay exponent $\gamma(\tau)$ defined as
\begin{align}
 \gamma(\tau) \equiv \frac{d \ln c^Z(\tau)}{d \ln \tau}
\end{align}
in Fig.~\ref{fig:3dkubojjrunningb}(c).
For $W=4$, the bottom of the running decay exponent $\gamma(\tau)$ is at $-0.8$, showing the
absence of a 3D behavior.
For $W=8$, the bottom drops to about
$-1.1$, showing a weak 3D behavior. For $W\geq16$, the
bottoms of the $\gamma(\tau)$ tend to saturate at a $W$-independent value
$-1.2$. As $W$ increases, the $\gamma(\tau)$ stay at this value
for a longer time. This indicates that $\gamma=-1.2$ is the
asymptotic decay exponent for a ``real'' 3D system. Since the decay exponent $\gamma=-1$ in 2D lattices, the threshold time $\tau_c(W)$ of the 3D-1D
crossover can thus be reasonably defined as $\gamma(\tau_c(W))=-1$.
It can be estimated that $\tau_c(8)\approx 35$ and $\tau_c(16)\approx 90$. For
$W\geq32$, it is hard to estimate the threshold time due to large
statistical errors.

\begin{figure}[ht]
\sidecaption[t]
\includegraphics[width=0.64\columnwidth]{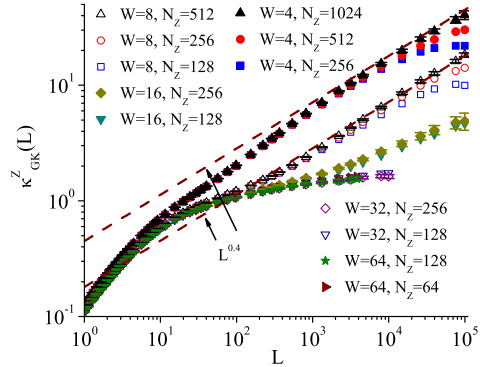}
\caption{\label{Fig:3DH4int} (color online).
$\kappa^Z_{\rm GK}(L)$ for various $W$ and $N_Z$.
For a given $W$, results for different values of $N_Z$ are plotted in
order to distinguish finite $N_Z$ effects. For each $W$
we plot error bars only for the data for the longest $N_Z$. One can
see the tendency of the curves to become flat as $W$ increases,
indicating the presence of normal heat conduction. Dashed lines with
slope 0.4 are drawn for reference.}
\end{figure}

The length-dependent thermal conductivity $\kappa^Z_{\rm GK}(L)$ for various $W$ and $N_Z$ is plotted in Fig.~\ref{Fig:3DH4int}.
similar to the 1D purely quartic lattice as shown in Fig.~\ref{fig:3dkubojjrunningb}(a) and (b), the 3D quartic lattice
$c(\tau)$ can be correctly calculated for quite large correlation time $\tau$ by simulating a not-very-long lattice $N_Z$, i.e. $L=v_s \tau >> N_Z$.
We are thus able to evaluate $\kappa^Z_{\rm GK}(L)$ for an effective length
$L$, which is much longer than $N_Z$. For $W=4$, the 3D behavior is nearly absent, similar to the picture
shown in Fig.~\ref{fig:3dkubojjrunningb}. As a result, the $\kappa^Z_{\rm GK}$ approaches to a 1D power-law divergence as $L^{0.4}$
directly. For $W=8$, beyond trivial ballistic regime,
the $\kappa^Z_{\rm GK}$ increases slowly at first, indicating the tendency
to 3D behavior, and then inflects up to the 1D-like power-law behavior. For
$W=16$, the inflection occurs at a larger length $L$. Finally for $W=32$
and $W=64$, although the similar inflection is expected to occur at
even larger length $L$, we are not able to see it due to numerical
difficulties.

One can conclude that the 3D system should display normal heat
conduction behavior if the cross-section area $W^2$ is large enough. Based on the threshold correlation time
$\tau_c(W)$ defined earlier, a threshold length $N^Z_c(W)$ can be defined
accordingly by requiring $N^Z_c(W)\equiv v_s \tau_c(W)$. The threshold length $N^Z_c(W)$
determined here is shorter than the estimation made by Saito and
Dhar~\cite{PhysRevLett.104.040601}, in which a lattice width $W=16$ shows a 3D
behavior up to $L=16384$. In a recent experimental study, an apparent 1D-like anomalous heat conduction behavior appears in
multiwall nanotubes with diameters around 10 nm and lengths of a few
$\mu$m~\cite{PhysRevLett.101.075903}. It seems that our estimation agrees
with the experimental result. However, more detailed experimental
measurements of heat conduction in shorter samples or samples with
larger cross-section of silicon nanowire \cite{Science.279.208,Science.299.1874}
or graphene \cite{NatMater.9.555}, are
necessary to give an accurate estimation of the threshold length $N^Z_c(W)$.

A running exponent $\alpha(L)$ is defined as the local slope of $\kappa^Z_{\rm GK}(L)$ as
\begin{equation} \label{alpha}
 \alpha(L) \equiv \frac{d \ln\kappa^Z_{\rm GK}(L)}{d \ln L}= \frac{L}{\kappa^Z_{\rm GK}(L)} c^Z(\tau).
\end{equation}
In the 3D regime, the $c^Z(\tau)$ behaves as
$c^Z(\tau) \sim \tau^{\gamma}$ as shown in Fig.~\ref{fig:3dkubojjrunningb}.
As $L$$\rightarrow$$\infty$, the $\kappa^Z_{\rm GK}(L)$
approaches a constant $\kappa$ since $\gamma<-1$. Then one can obtain
\begin{equation}
 \alpha(L)  \sim \frac{L}{\kappa} L^{\gamma}=\frac{1}{\kappa} L^{\gamma+1}.
\end{equation}
where $\alpha(L)$ decays asymptotically as $L^{-0.2}$ for $\gamma=-1.2$.
This power-law decay of the exponent $\alpha(L)$ with the length $L$ as $\alpha(L)\propto L^{-0.2}$ quantitatively explains the result previously found in Ref.~\cite{PhysRevLett.104.040601}.

In summary, we have numerically studied heat conduction in
3D momentum-conserving nonlinear lattices by the
Green-Kubo method. The main findings are: (1) For a fixed
width $W\geq8$, a 3D-1D crossover was found to occur at a
$W$-dependent threshold of a lattice size $N^Z_c(W)$. Below $N^Z_c$
the system displays a 3D behavior while it displays a
1D behavior above $N^Z_c$. (2) In the 3D regime, the heat current autocorrelation
function $c^Z(\tau)$ decays asymptotically as $\tau^{\gamma}$ with
${\gamma}=-1.2$. This value being more negative than $-1$ indicates normal heat conduction,
which is consistent with the theoretical expectation.
(3) The exponent $\gamma=-1.2$ implies that the running exponent $\alpha(L)$ follows a
power-law decay, $\alpha\propto L^{-0.2}$, which also agrees very well
with that reported in Ref.~\cite{PhysRevLett.104.040601}. (4) The detailed value
$\gamma=-1.2$ however deviates significantly from the conventional theoretical expectation of $\gamma=-1.5$.

\section{Simulation of heat transport with the diffusion method}
In the numerical studies of heat transport in nonlinear lattices, the most frequently used methods are the direct non-equilibrium heat bath method~\cite{PhysRep.377.1} and the equilibrium Green-Kubo method~\cite{GREENKUBO}. For the non-equilibrium heat bath method, the system is connected with two heat baths in both ends and driven into a stationary state. The averaged heat flux $j$ is recorded which gives rise to the thermal conductivity $\kappa$ through the relation of $j=-\kappa \nabla{T}$. For the equilibrium Green-Kubo method, the system is prepared from microcanonical dynamics without heat bath. The autocorrelation function $C_{JJ}(t)$ of the total heat flux is recorded and the thermal conductivity $\kappa$ can be obtained by integrating $C_{JJ}(t)$ via the Green-Kubo formula.

Besides the non-equilibrium heat bath method and Green-Kubo method, a novel diffusion method is recently proposed by Zhao in studying the anomalous heat transport and diffusion processes of 1D nonlinear lattices~\cite{PhysRevLett.96.140602}. This is also an equilibrium method, while the statistics can be drawn from microcanonical or canonical dynamics. In contrast to the Green-Kubo method, this diffusion method relies on the information of the autocorrelation function of the local energy. In Hamiltonian dynamics, the total energy is always a conserved quantity. Due to this very property of energy conservation, it is then rigorously proved in a heat diffusion theory recently developed by Liu {\it et al}, stating that the energy diffusion method is equivalent to the Green-Kubo method in the sense of determining the system's thermal conductivity~\cite{PhysRevLett.112.040601}. In particular, the energy diffusion method is able to provide more information than that from the Green-Kubo method. The real-time spatiotemporal excess energy density distribution $\rho_E(x,t)$ plays the role of a generating function which is essential for the analysis of underlying heat conduction behavior.

In principle, the thermal conductivity $\kappa$ can be generally expressed in a length-dependent form as $\kappa\propto L^{\alpha}$ with $L$ the system length. For system with normal heat conduction, the heat divergency exponent $\alpha=0$ implies that $\kappa$ is length-independent obeying the Fourier's heat conduction law. The heat divergency exponent $\alpha=1$ represents a ballistic heat conduction behavior. For $0<\alpha<1$, the system displays the so-called anomalous heat conduction behavior. On the other hand, the Mean Square Displacement (MSD) $\langle\Delta{x^2(t)}\rangle_{\mbox{E}}$ of the excess energy generally grows with time asymptotically as $\langle\Delta{x^2(t)}\rangle_{\mbox{E}}\propto t^{\beta}$ where the energy diffusion exponent $\beta$ classifies the diffusion behaviors. The $\beta=1$ and $2$ represent the normal and ballistic energy diffusion behaviors, respectively. For $1<\beta<2$, the diffusion process is called superdiffusion.

In 1D nonlinear lattice systems, the heat conduction and energy diffusion originating from the same energy transport process are closely related. What is the relation between the heat conduction and energy diffusion processes? The relation formula between heat divergency and energy diffusion exponents $\alpha=\beta-1$ is proposed by Cipriani {\it et. al} by investigating single particle Levy walk diffusion process~\cite{PhysRevLett.94.244301}. This same formula is then formally derived as a natural result from the heat diffusion theory~\cite{PhysRevLett.112.040601}. This relation formula tells that: (i) Normal energy diffusion with $\beta=1$ corresponds to normal heat conduction with $\alpha=0$ and vice versa. This is the case for 1D $\phi^4$ lattice and Frenkel-Kontorova lattice~\cite{PhysRevLett.96.140602}. (ii) Ballistic energy diffusion with $\beta=2$ implies ballistic heat conduction with $\alpha=1$ and vice versa. The 1D Harmonic lattice and Toda lattice fall into this class~\cite{PhysRevLett.96.140602}. (iii) Superdiffusive energy diffusion with $1<\beta<2$ yields anomalous heat conduction with $0<\alpha<1$ and vice versa. The 1D FPU-$\beta$ lattice is verified to posses energy superdiffusion with $\beta=1.40$ and anomalous heat conduction with $\alpha=0.40$~\cite{PhysRevLett.96.140602} belonging to this class.

Besides total energy, total momentum is another conserved quantity for 1D nonlinear lattices without on-site potential, such as the FPU-$\beta$ lattice. It is commonly believed that the conservation of momentum is essential for the actual heat conduction behavior. Predictions from mode coupling theory~\cite{PhysRep.377.1} and renormalization group theory~\cite{PhysRevLett.89.200601} claim that momentum conservation should give rise to anomalous heat conduction in one dimensional systems. However, there is one exception to these predictions: the 1D coupled rotator lattice, which displays normal heat conduction behavior despite its momentum conserving nature~\cite{PhysRevLett.84.2381,PhysRevLett.84.2144}. This unusual phenomenon stimulates the efforts to explore the interplay between energy transport and momentum transport. The transport coefficient corresponding to the momentum transport is the bulk viscosity. For momentum conserving system, in principle, there should also be a formal connection between the momentum transport and the momentum diffusion. This very momentum diffusion theory has also been developed for 1D momentum-conserving lattices~\cite{LI_NJP}. Due to the complexity of bulk viscosity, the momentum diffusion theory is more complicated than the heat diffusion theory. Nevertheless, there seems to be a relation between the actual behaviors of energy and momentum transport implied from extensive numerical studies.

In the following, the energy diffusion method will be first introduced. The heat diffusion theory will be derived in the framework of linear response theory. Numerical simulations for two typical 1D nonlinear lattices will be used to verify the validity of this heat diffusion theory. Some results from the energy diffusion method will be shown for typical 1D lattices. The momentum diffusion method will then be discussed. The momentum diffusion theory will be derived in the same sense as heat diffusion theory for the 1D lattice systems. Some numerical results from the momentum diffusion method will be displayed and potential connection between momentum and heat transports will be discussed in the final part.

\subsection{Energy diffusion}
In this part, we first introduce the energy diffusion method in the investigation for energy diffusion process of 1D lattices. The heat diffusion theory will be derived in the linear response regime and verified by numerical simulations. Some numerical results from this energy diffusion method will be presented to demonstrate the advantages of this novel method.

\subsubsection{Heat diffusion theory}
The heat diffusion theory~\cite{PhysRevLett.112.040601,LI_NJP} unifies energy diffusion and heat conduction in a rigorous way. The central result reads
\begin{equation}\label{central-result}
\frac{d^2\langle\Delta x^2(t)\rangle_{\mbox{E}}}{dt^2}=\frac{2C_{JJ}(t)}{k_B T^2 c_{\mbox{v}}} \;,
\end{equation}
where $k_B$ is the Boltzmann constant and $c_{\mbox{v}}$ is the volumetric specific heat. The autocorrelation of total heat flux $C_{JJ}(t)$ on the right hand side is the term entering the Green-Kubo formula from which thermal conductivity can be calculated. The MSD $\langle\Delta x^2(t)\rangle_{\mbox{E}}$ of energy diffusion describes the relaxation process in which an initially nonequilibrium energy distribution evolves towards equilibrium:
\begin{equation}\label{msd}
\langle\Delta x^2(t)\rangle_{\mbox{E}}\equiv \int (x-\langle x\rangle_E)^2\rho_E(x,t)dx=\langle x^2(t)\rangle_E-\langle x\rangle^2_E\;.
\end{equation}
This normalized fraction of excess energy $\rho_E(x,t)$ at a certain position $x$ at time $t$ reads
\begin{equation}\label{eq-nb:rhoe1}
\rho_E(x,t)=\frac{\delta\langle h(x,t)\rangle_{\mbox{neq}}}{\delta E}=\frac{\delta\langle h(x,t)\rangle_{\mbox{neq}}}{\int\delta\langle h(x,0)\rangle_{\mbox{neq}}dx}.
\end{equation}
Here the excess energy distribution is proportional to the deviation $\delta\langle h(x,t)\rangle_{\mbox{neq}}\equiv\langle h(x,t)\rangle_{\mbox{neq}}-\langle h(x)\rangle_{\mbox{eq}}$, where $\langle\cdot\rangle_{\mbox{neq}}$ denotes the expectation value in the nonequilibrium diffusion process, $\langle\cdot\rangle_{\mbox{eq}}$ denotes the equilibrium average, and $h(x,t)$ denotes the local Hamiltonian density. For isolated energy-conserving systems, this total excess energy, $\delta E=\int \delta\langle h(x,t)\rangle_{\mbox{neq}}dx$ remains conserved. Therefore, the normalized condition $\int\rho_E(x,t)dx=1$ is fulfilled during the time evolution as a result of energy conservation.

In the linear response regime, the deviation of local excess energy can be explicitly derived in terms of equilibrium spatiotemporal correlation $C_{hh}(x,t;x',0)$ of local Hamiltonian density $h(x,t)$ as
\begin{equation}\label{eq-nb:rhoe2}
\delta \langle h(x,t)\rangle_{\mbox{neq}}=\frac{1}{k_B T}\int C_{hh}(x,t;x',0)\eta(x')dx',
\end{equation}
where $C_{hh}(x,t;x',t')\equiv \langle\Delta h(x,t)\Delta h(x',t')\rangle_{\mbox{eq}}$, with $\Delta h(x,t)=h(x,t)-\langle h(x)\rangle_{\mbox{eq}}$, and the $-\eta(x)h(x)$ represents a small perturbation switched off suddenly at time $t=0$, with $\eta(x)\ll 1$.

Therefore, the normalized excess energy distribution can be derived from Eq. (\ref{eq-nb:rhoe1}) and (\ref{eq-nb:rhoe2}) as
\begin{equation}\label{rho-E}
\rho_E(x,t)=\frac{1}{\mathscr{N}}\int C_{hh}(x-x',t)\eta(x')dx',
\end{equation}
where $\mathscr{N}=k_B T^2 c_{\mbox{v}}\int \eta(x)dx$ is the normalization constant.

The key point which connects energy diffusion and heat conduction is the local energy continuity equation due to energy conservation
\begin{equation}
\frac{\partial{h(x,t)}}{\partial{t}}+\frac{\partial{j(x,t)}}{\partial{x}}=0,
\end{equation}
where $j(x,t)$ is the local heat flux density. One can then obtain
\begin{equation}
\frac{\partial^2{C_{hh}(x,t)}}{\partial{t^2}}=\frac{\partial^2{C_{jj}(x,t)}}{\partial{x^2}},
\end{equation}
By defining the total heat flux $J_L=\int^{L/2}_{-L/2}j(x,t)dx$ and the autocorrelation function of total heat flux $C_{JJ}(t)\equiv\lim_{L\rightarrow\infty}\langle J_L(t)J_L(0)\rangle_{\mbox{eq}}/L=\int^{\infty}_{-\infty}C_{jj}(x,t)dx$, the central result (\ref{central-result}) of the heat diffusion theory can be derived.

As a result of energy conservation, the heat diffusion theory of Eq. (\ref{central-result}) gives the general relation between energy diffusion and heat conduction. The actual behavior of energy diffusion or heat conduction can be normal or anomalous while the relation (\ref{central-result}) remains to be the same:

(I) For normal energy diffusion, the MSD increases asymptotically linearly with time, i.e.
\begin{equation}
\langle \Delta x^2(t)\rangle_{\mbox{E}}\cong 2D_E t,
\end{equation}
in the infinite time limit $t\rightarrow\infty$. Here $D_E$ is the so-called thermal diffusivity. According to Eq. (\ref{central-result}), the corresponding thermal conductivity $\kappa$ can be obtained as
\begin{equation}
\kappa=\int^{\infty}_{0}\frac{C_{JJ}(t)}{k_{B}T^2}dt=\frac{c_{\mbox{v}}}{2}\lim_{t\rightarrow\infty}\frac{d\langle \Delta x^2(t)\rangle_{\mbox{E}}}{dt}=c_{\mbox{v}}D_E.
\end{equation}
This is nothing but the Green-Kubo expression for normal heat conduction.

(II) For ballistic energy diffusion, the MSD is asymptotically proportional to the square of time as
\begin{equation}
\langle \Delta x^2(t)\rangle_{\mbox{E}} \propto t^2.
\end{equation}
Substituting this expression into Eq. (\ref{central-result}), one can deduce that $C_{JJ}(t)$ is a non-decaying constant, reflecting the ballistic nature of heat conduction as well.

(III) For superdiffusive energy diffusion, the MSD obeys
\begin{equation}
\langle \Delta x^2(t)\rangle_{\mbox{E}} \propto t^{\beta},\,\,\,1<\beta<2.
\end{equation}
From Eq. (\ref{central-result}), the decay of $C_{JJ}(t)$ is a slow process as $C_{JJ}(t)\propto t^{\beta-2}$ and the integral of $C_{JJ}(t)$ diverges. In this situation, no finite superdiffusive thermal conductivity exists. The typical way in practice is to introduce an upper cutoff time $t_s\sim L/v_s$ with $v_s$ the speed of sound due to renormalized phonons~\cite{PhysRevLett.105.054102}. A length-dependent superdiffusive thermal conductivity can be obtained through Eq. (\ref{central-result}):
\begin{equation}
\kappa\sim \frac{1}{k_{B}T^2}\int^{L/v_s}_{0}C_{JJ}(t)dt=\frac{c_{\mbox{v}}}{2}\left.\frac{d\langle \Delta x^2(t)\rangle_{\mbox{E}}}{dt}\right\rvert_{t\sim L/v_s}\propto L^{\beta-1}.
\end{equation}
The length-dependent anomalous thermal conductivity is usually expressed as $\kappa\propto L^{\alpha}$. One can immediately obtain the scaling relation between energy diffusion and heat conduction
\begin{equation}\label{heat-diffusion-formula}
\alpha=\beta-1,
\end{equation}
which is a general relation and not limited to superdiffusive energy diffusion only.

(IV) For subdiffusive energy diffusion, the MSD follows asymptotically
\begin{equation}
\langle \Delta x^2(t)\rangle_{\mbox{E}} \propto t^{\beta},\,\,\,0<\beta<1.
\end{equation}
From the relation in (\ref{central-result}), the autocorrelation function of total heat flux reads asymptotically
\begin{equation}\label{prefactor-subdif}
C_{JJ}(t)\propto \beta(\beta-1)t^{\beta-2}.
\end{equation}
The $C_{JJ}(t)$ remains integrable and the thermal conductivity can be derived as
\begin{equation}
\kappa=\int^{\infty}_{0}\frac{C_{JJ}(t)}{k_{B}T^2}dt=\frac{c_{\mbox{v}}}{2}\lim_{t\rightarrow\infty}\frac{d\langle \Delta x^2(t)\rangle_{\mbox{E}}}{dt}\sim \lim_{t\rightarrow\infty}t^{\beta-1}=0.
\end{equation}
This vanishing integral of $C_{JJ}(t)$ is not surprising, if one notices that the asymptotic prefactor of $C_{JJ}(t)$ in (\ref{prefactor-subdif}) is a negative value due to $\beta-1<0$.

\subsubsection{Numerical verification of the heat diffusion theory}

\begin{figure}[t]
\sidecaption
\includegraphics[scale=.4]{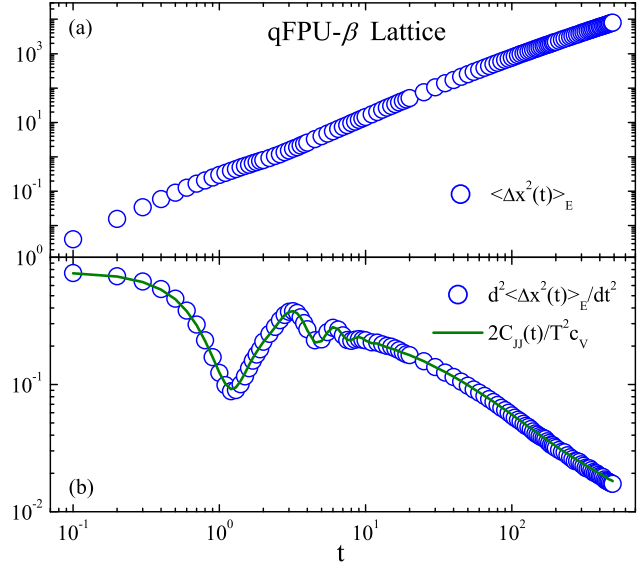}
%
%
\caption{Numerical verification of the main relation of Eq. (\ref{central-result}) of heat diffusion theory for 1D qFPU-$\beta$ lattice. (a) The MSD $\langle \Delta x^2(t)\rangle_{\mbox{E}}$ of energy diffusion as the function of time $t$ in log-log scale. (b) The second derivative of MSD $d^2\langle \Delta x^2(t)\rangle_{\mbox{E}}/dt^2$ (hollow circles) and the rescaled autocorrelation function of total heat flux $2C_{JJ}(t)/(T^2c_{\mbox{v}})$ (solid line) as the function of time $t$ in log-log scale. The perfect agreement between them demonstrates the validity of the main relation of Eq. (\ref{central-result}). The Boltzmann constant $k_B$ has been set as unity when applying dimensionless units in numerical simulations. The simulations are performed for a qFPU-$\beta$ lattice with the average energy density per atom $E=0.015$ and the number of atoms $N=601$.}
\label{fig:Nianbei-qfpub-compare}       
\end{figure}

The heat diffusion theory of Eq. (\ref{central-result}) developed for continuous system is general and applicable also for discrete system. We choose two typical 1D nonlinear lattices to demonstrate the validity of this heat diffusion theory. One is the purely quartic FPU-$\beta$ (qFPU-$\beta$) lattice which is the high temperature limit of FPU-$\beta$ lattice, where energy diffusion is superdiffusive and heat conduction is anomalous or length-dependent~\cite{PhysRevLett.98.184301,EurophysLett.93.54002}. The dimensionless Hamiltonian with finite $N=2M+1$ atoms reads
\begin{equation}
H=\sum_i H_i=\sum_i\left[\frac{1}{2}p^2_i+\frac{1}{4}(u_{i+1}-u_i)^4\right],
\end{equation}
where $p_i$ and $u_i$ denote momentum and displacement from equilibrium position for $i$-th atom, respectively. The index $i$ is numerated from $-M$ to $M$.

The other one is the $\phi^4$ lattice which shows normal energy diffusion as well as normal heat conduction~\cite{PhysRevLett.96.140602,PhysRevE.61.3828,Aoki2000250}. The dimensionless Hamiltonian reads
\begin{equation}
H=\sum_i H_i=\sum_i\left[\frac{1}{2}p^2_i+\frac{1}{2}(u_{i+1}-u_i)^2+\frac{1}{4}u^4_i\right].
\end{equation}

In the numerical simulations, we adopt the microcanonical dynamics where energy density $E$ per atom is set as the input parameter. Algorithm with higher accuracy of fourth-order symplectic method~\cite{PhysRevE.79.056211,fourthorder} can be used to integrate the equations of motions. Periodic boundary conditions $u_i=u_{i+N}$ and $p_i=p_{i+N}$ are applied and the equilibrium temperature $T$ can be calculated from the definition $T=T_i=\langle p^2_i\rangle$, where $\langle \cdot\rangle$ denotes time average which equals to the ensemble average due to the chaotic and ergodic nature of these two nonlinear lattices. The volumetric specific heat need to be calculated via the relation $c_{\mbox{v}}=(\langle H^2_i\rangle-\langle H_i\rangle^2)/T^2$ which is independent of the choice of index $i$. For qFPU-$\beta$ lattice, the volumetric specific heat $c_{\mbox{v}}=0.75$ is a temperature-independent constant.

In order to define the discrete expression of excess energy density distribution $\rho_E(i,t)$, we first introduce the energy-energy correlation function, reading:
\begin{equation}
C_E(i,t;j,t=0)\equiv\frac{\langle \Delta H_i(t)\Delta H_j(0)\rangle}{k_{B}T^2c_{\mbox{v}}},
\end{equation}
where $\Delta H_i(t)\equiv H_i(t)-\langle H_i(t)\rangle$. Applying a localized, small initial excess energy perturbation at the central site, $\eta(i)=\varepsilon\delta_{i,0}$ in Eq. (\ref{rho-E}), the discrete excess energy distribution can be obtained:
\begin{equation}
\rho_E(i,t)=\sum_j C_E(i,t;j,0)\eta(j)/\varepsilon=C_E(i,t:j=0,t=0),\,\,-M\le i \le M.
\end{equation}
The MSD $\langle \Delta x^2(t)\rangle_{\mbox{E}}$ of energy diffusion of Eq. (\ref{msd}) for discrete lattice can be defined as
\begin{equation}
\langle \Delta x^2(t)\rangle_{\mbox{E}}\equiv\sum_i i^2\rho_E(i,t)=\sum_i i^2C_E(i,t;j=0,t=0),\,\,-M\le i \le M,
\end{equation}
by noticing that $\langle x(t)\rangle_{\mbox{E}}=0$.

The second derivative of $\langle \Delta x^2(t)\rangle_{\mbox{E}}$ can be numerically obtained as
\begin{equation}
\frac{d^2\langle \Delta x^2(t)\rangle_{\mbox{E}}}{dt^2}\approx\frac{\langle \Delta x^2(t+\Delta t)\rangle_{\mbox{E}}-2\langle \Delta x^2(t)\rangle_{\mbox{E}}+\langle \Delta x^2(t-\Delta t)\rangle_{\mbox{E}}}{(\Delta t)^2}
\end{equation}
where $\Delta t$ is the time difference between two consecutive recorded $\langle \Delta x^2(t)\rangle_{\mbox{E}}$. The autocorrelation function of total heat flux $C_{JJ}(t)$ for discrete lattice system is defined as $C_{JJ}(t)=\langle \Delta J(t)\Delta J(0)\rangle/N$, with $J(t)=\sum_i j_i(t)$. The local heat flux $j_i(t)=-\dot{u}_i\partial{V(u_i-u_{i-1})}/\partial{u_i}$ is derived from local energy continuity equation where $V(x)$ denotes the form of potential energy in Hamiltonian.

\begin{figure}[t]
\sidecaption
\includegraphics[scale=.5]{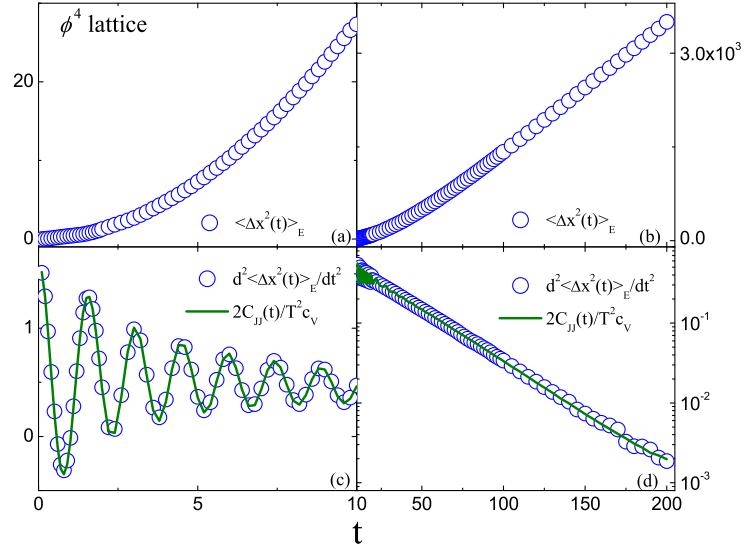}
%
%
\caption{Numerical verification of the main relation of Eq. (\ref{central-result}) of heat diffusion theory for 1D $\phi^4$ lattice. (a) and (b): The MSD $\langle \Delta x^2(t)\rangle_{\mbox{E}}$ of energy diffusion as the function of time $t$ for $0<t<10$ and $10<t<200$, respectively. (c) and (d): The second derivative of MSD $d^2\langle \Delta x^2(t)\rangle_{\mbox{E}}/dt^2$ (hollow circles) and the rescaled autocorrelation function of total heat flux $2C_{JJ}(t)/(T^2c_{\mbox{v}})$ (solid line) as the function of time $t$ in linear-linear scale for $0<t<10$ and log-linear scale for $10<t<200$, respectively. The perfect agreement between them demonstrates the validity of the main relation of Eq. (\ref{central-result}). The simulations are performed for a $\phi^4$ lattice with the average energy density per atom $E=0.4$ and the number of atoms $N=501$.}
\label{fig:Nianbei-phi4-compare}       
\end{figure}

In Fig. \ref{fig:Nianbei-qfpub-compare}, we verify the main relation (\ref{central-result}) for 1D qFPU-$\beta$ lattice. The MSD of energy diffusion $\langle \Delta x^2(t)\rangle_{\mbox{E}}$ as the function of time is plotted in Fig. \ref{fig:Nianbei-qfpub-compare} (a). Its second derivative $d^2\langle \Delta x^2(t)\rangle_{\mbox{E}}/dt^2$ is extracted out and directly compared with the rescaled autocorrelation function of total heat flux $2C_{JJ}(t)/(k_BT^2c_{\mbox{v}})$ in Fig. \ref{fig:Nianbei-qfpub-compare} (b). The perfect agreement between them justifies the validity of the main relation (\ref{central-result}) predicted from heat diffusion theory. The numerical results for 1D $\phi^4$ lattice are also plotted in Fig. \ref{fig:Nianbei-phi4-compare} and same conclusion can be obtained. It should be pointed out that the 1D qFPU-$\beta$ lattice displays superdiffusive energy diffusion and anomalous heat conduction, while 1D $\phi^4$ lattice shows normal energy diffusion and heat conduction. These facts can be observed by noticing that the autocorrelation function $C_{JJ}(t)$ eventually follows a power law decay as $C_{JJ}(t)\propto t^{-0.60}$ for qFPU-$\beta$ lattice and an exponential decay as $C_{JJ}(t)\propto e^{-t/\tau}$ for $\phi^4$ lattice where $\tau$ represents a characteristic relaxation time.

\subsubsection{Energy diffusion properties for typical 1D lattices}

\begin{figure}[t]
\sidecaption
\includegraphics[scale=.5]{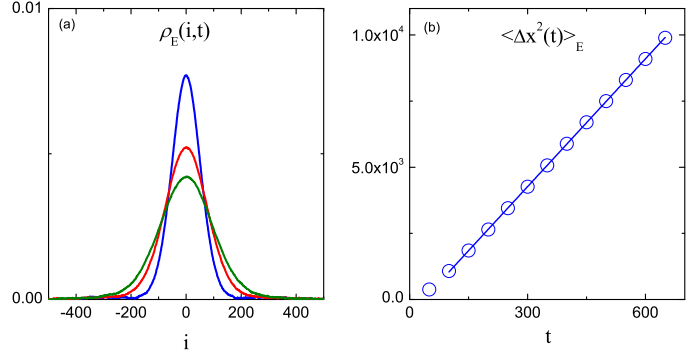}
%
%
\caption{Energy diffusion processes in the 1D coupled rotator lattice. (a) Spatial distribution of the energy autocorrelation $\rho_E(i,t)=C_E(i,t;j=0,t=0)$. The correlation times are $t=200, 400$ and $600$ from top to the bottom in the central part, respectively. The distirbution of $\rho_E(i,t)$ follows the Gaussian normal distribution as $\rho_E(i,t)\sim\frac{1}{\sqrt{4\pi D_E t}}e^{-\frac{i^2}{4D_E t}}$. (b) The MSD of the energy diffusion $\langle \Delta x^2(t)\rangle_{\mbox{E}}$ as the function of time. The solid straight line is the best fit for the MSD $\langle \Delta x^2(t)\rangle_{\mbox{E}}$ implying a normal diffusion process. The simulations are performed for a coupled rotator lattice with the average energy density per atom $E=0.45$ and the number of atoms $N=1501$.}
\label{fig:Nianbei-rotator-energy}       
\end{figure}

The heat diffusion theory formally connects the energy diffusion and heat conduction as described in Eq. (\ref{central-result}). It enables us to use energy diffusion method to investigate the heat conduction process. More importantly, the energy diffusion method is able to provide more information about the heat transport process than the Green-Kubo method or the direct non-equilibrium heat bath method.

The key information from the energy diffusion method is the spatiotemporal distribution of excess energy $\rho_E(i,t)$, from which the MSD of energy diffusion $\langle \Delta x^2(t)\rangle_{\mbox{E}}$ can be generated. The second derivative of $\langle \Delta x^2(t)\rangle_{\mbox{E}}$ gives rise to the autocorrelation function of total heat flux $C_{JJ}(t)$ which finally yields the thermal conductivity via Green-Kubo formula. With the knowledge of $\rho_E(i,t)$, one can resolve the expression of $\langle \Delta x^2(t)\rangle_{\mbox{E}}$ or $C_{JJ}(t)$, but not vice versa. Therefore, in determining the actual heat conduction behavior, the excess energy distribution $\rho_E(i,t)$ plays the essential role of a generating function.

To illustrate the importance \cite{LI_NJP}: the coupled rotator lattice
\begin{equation}\label{cr-ham}
H=\sum_i\left[\frac{1}{2}p^2_i+(1-\cos{(u_{i+1}-u_i)})\right],
\end{equation}
and the amended coupled rotator lattice
\begin{equation}\label{ar-ham}
H=\sum_i\left[\frac{1}{2}p^2_i+(1-\cos{(u_{i+1}-u_i)})+\frac{K}{2}(u_{i+1}-u_i)^2\right],
\end{equation}
where an additional quadratic interaction potential term is added.

\begin{figure}[t]
\sidecaption
\includegraphics[scale=.5]{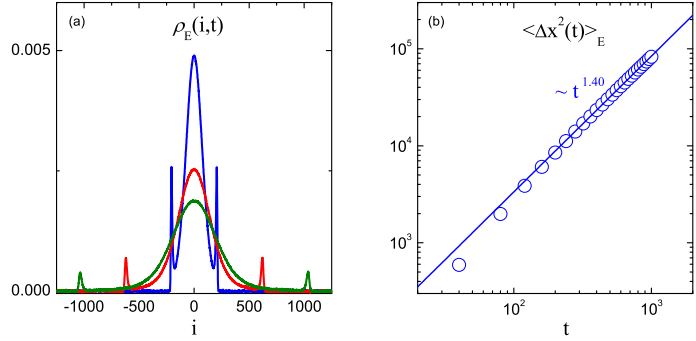}
%
%
\caption{Energy diffusion processes in the 1D amended coupled rotator lattice. (a) Spatial distribution of the energy autocorrelation $\rho_E(i,t)=C_E(i,t;j=0,t=0)$. The correlation times are $t=200, 600$ and $1000$ from top to the bottom in the central part, respectively. Besides the central peak, there are two side peaks moving outside with the constant sound velocity. This is the Levy walk distribution giving rise to a superdiffusive energy diffusion. (b) The MSD of the energy diffusion $\langle \Delta x^2(t)\rangle_{\mbox{E}}$ as the function of time. The solid straight line is the best fit for the superdiffusive MSD as $\langle \Delta x^2(t)\rangle_{\mbox{E}}\propto t^{1.40}$. The simulations are performed for an amended coupled rotator lattice with the average energy density per atom $E=1$. The number of atoms $N=2501$ and $K=0.5$.}
\label{fig:Nianbei-amendedrotator-energy}       
\end{figure}

The excess energy distributions $\rho_E(i,t)$ for coupled rotator lattice are plotted in Fig. \ref{fig:Nianbei-rotator-energy} (a). For sufficiently large times, the excess energy distribution $\rho_E(i,t)$ evolves very closely into a Gaussian distribution function with its profile perfectly given by
\begin{equation}
\rho_E(i,t)\sim\frac{1}{\sqrt{4\pi D_E t}}e^{-\frac{i^2}{4D_E t}},
\end{equation}
where $D_E$ denotes the diffusion constant for energy diffusion. As a result, the MSD of energy diffusion $\langle \Delta x^2(t)\rangle_{\mbox{E}}$ depicts at a linear time dependence
\begin{equation}
\langle \Delta x^2(t)\rangle_{\mbox{E}}\sim \sum_i i^2\rho_E(i,t)=\sum_i i^2\frac{1}{\sqrt{4\pi D_E t}}e^{-\frac{i^2}{4D_E t}}=2D_Et,
\end{equation}
for sufficiently long time as can be seen in Fig. \ref{fig:Nianbei-rotator-energy} (b), being the hall mark for normal diffusion. Accordingly, heat diffusion theory for normal energy diffusion implies that the heat conduction behavior is normal as well, with the thermal conductivity given by $\kappa=c_{\mbox{v}}D_E$.

This normal energy diffusion behavior also occurs for other 1D lattice systems with normal heat conduction, such as $\phi^4$ lattice and Frenkel-Kontorova lattice~\cite{PhysRevLett.96.140602}.

For the 1D amended coupled rotator lattice described in Eq. (\ref{ar-ham}), the excess energy distributions $\rho_E(i,t)$ are plotted in Fig. \ref{fig:Nianbei-amendedrotator-energy} (a). Besides the central peak, there are also two side peaks moving outside with a constant sound velocity $v_s$. It is amazing that this excess energy distribution $\rho_E(i,t)$ of 1D nonlinear lattice closely resembles to the single particle Levy walk distribution $\rho_{LW}(x,t)$~\cite{PhysRevLett.94.244301}
\begin{equation}\label{Nianbei-levy-walk}
\rho_{LW}(x,t)\propto\left\{
\begin{array}{ll}
t^{-1/\mu}\exp{(\frac{-ax^2}{t^{2/\mu}})},\,\, & |x|\lesssim t^{1/\mu}\\
tx^{-\mu-1},\,\, & t^{1/\mu}<|x|<vt\\
t^{1-\mu},\,\, & |x|=vt\\
0,\,\, & |x|>vt
\end{array}
\right.
\end{equation}
where $a$ is an unknown constant and $v$ is the particle velocity. This Levy walk distribution $\rho_{LW}(x,t)$ is a result of a particle moving ballistically between consecutive collisions with a waiting time distribution $\psi(t)\propto t^{-\mu-1}$ and a velocity distribution $f(u)=[\delta(u-v)+\delta(u+v)]/2$.

The MSD for the Levy walk distribution $\rho_{LW}(x,t)$ follows a time dependence of $\langle \Delta x^2(t)\rangle_{LW}\propto t^{\beta}$ with $\beta=3-\mu$. In Fig. \ref{fig:Nianbei-amendedrotator-energy} (b), the time dependence of MSD $\langle \Delta x^2(t)\rangle_{\mbox{E}}$ of energy diffusion for the 1D amended coupled rotator lattice is plotted where the best fit indicates that $\beta=1.40$. This will in turn correspond to a $\mu=1.60$ in the Levy walk scenario. According to the relation formula of Eq. (\ref{heat-diffusion-formula}) of heat diffusion theory, the corresponding heat conduction should be anomalous with a divergent length dependent thermal conductivity of $\kappa\propto L^{\alpha}$ with $\alpha=0.40$~\cite{EurophysLett.93.54002}.

It is very interesting to notice that there is a characteristic $\mu$ for the Levy walk distribution. By requiring that the heights of the central peak and side peaks decay with a same rate in the Levy walk distribution (\ref{Nianbei-levy-walk}), one can obtain $-1/\mu=1-\mu$ which gives rise to the golden ratio $\mu=(\sqrt{5}+1)/2\approx 1.618$. As a result, the corresponding characteristic energy supperdiffusion exponent $\beta=(5-\sqrt{5})/2\approx 1.382$ and the anomalous heat conduction exponent $\alpha=(3-\sqrt{5})/2\approx 0.382$ can be derived. Interesting enough, Lee-Dadswell \textit{et al.} derived the same exponent $\alpha=(3-\sqrt{5})/2$ as the converging value of a Fibonacci sequence in a toy model within the framework of hydrodynamical theory in 2005 \cite{LeeDadswell2005}.  Actually, this exponent is not far from the existing numerical results~\cite{PhysRevLett.98.184301,EurophysLett.93.54002,PhysRevLett.94.025507}.

\subsection{Momentum diffusion}
In the following part, the momentum diffusion method will be introduced for momentum conserving systems. A momentum diffusion theory will be derived in the same sense as heat diffusion theory. The numerical results reflecting the momentum diffusion properties will then be presented for several 1D nonlinear lattices. Based on the numerical results, the possible connection between momentum and energy transports will be discussed in the final part.

\subsubsection{Momentum diffusion theory}
In analogy to the heat diffusion, one can also construct a momentum diffusion theory~\cite{LI_NJP} for lattice systems which reads:
\begin{equation}\label{momentum-diffusion-theory}
\frac{d^2\langle \Delta x^2(t)\rangle_{\mbox{P}}}{dt^2}=\frac{2}{k_{B}T}C_{J^{P}}(t),
\end{equation}
where $\langle \Delta x^2(t)\rangle_{\mbox{P}}$ denotes the momentum diffusion and $C_{J^P}(t)$ is the centered autocorrelation function of total momentum flux.

\begin{figure}[t]
\sidecaption
\includegraphics[scale=.5]{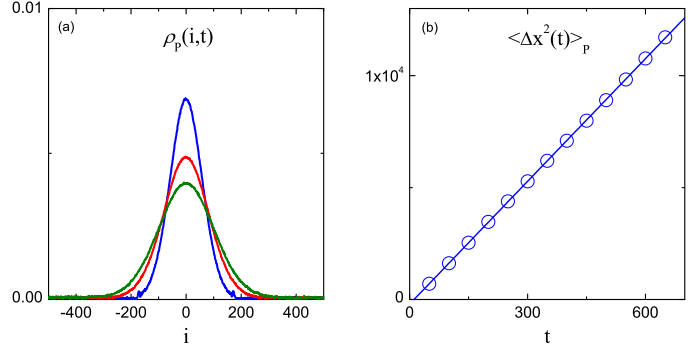}
%
%
\caption{Momentum diffusion processes in the 1D coupled rotator lattice. (a) Spatial distribution of the energy autocorrelation $\rho_P(i,t)=C_P(i,t;j=0,t=0)$. The correlation times are $t=200, 400$ and $600$ from top to the bottom in the central part, respectively. The distirbution of $\rho_P(i,t)$ follows the Gaussian normal distribution as $\rho_P(i,t)\sim\frac{1}{\sqrt{4\pi D_P t}}e^{-\frac{i^2}{4D_P t}}$. (b) The MSD of the momentum diffusion $\langle \Delta x^2(t)\rangle_{\mbox{P}}$ as the function of time. The solid straight line is the best fit for the MSD $\langle \Delta x^2(t)\rangle_{\mbox{P}}$ implying a normal diffusion process. The simulations are performed for a coupled rotator lattice with the average energy density per atom $E=0.45$ and the number of atoms $N=1501$.}
\label{fig:Nianbei-rotator-momentum}       
\end{figure}

The MSD of excess momentum $\langle \Delta x^2(t)\rangle_{\mbox{P}}$ can be defined as
\begin{equation}
\langle \Delta x^2(t)\rangle_{\mbox{P}}=\sum_i i^2\rho_{P}(i,t),\,\,-M\le i\le M.
\end{equation}
The excess momentum distribution $\rho_{P}(i,t)$ describes the nonequilibrium relaxation process of momentum due to a small kick of short duration to the $j$-th atom. The kick occurs with a constant impulse $I$, yielding a force kick at site $j$ as $f_j(t)=I\delta(t)$. The normalized $\rho_{P}(i,t)$ is given by
\begin{equation}
\rho_{P}(i,t)=\frac{\langle p_i(t)\rangle_{\mbox{re}}}{\sum_i\langle p_i(t)\rangle_{\mbox{re}}},
\end{equation}
where $\langle p_i(t)\rangle_{\mbox{re}}$ represents the response of momentum of $i$-th atom to the small perturbation of $-f_j(t)u_j$. In the linear response regime, it can be obtained that $\langle p_i(t)\rangle_{\mbox{re}}=IC_P(i,t;j,0)$, where
\begin{equation}
C_P(i,t;j,0)=\frac{\langle \Delta p_i(t)\Delta p_j(0)\rangle}{k_{B}T}
\end{equation}
is the autocorrelation function for the excess momentum fluctuation. The sum $\sum_iC_P(i,t;j,0)=1$ at time $t=0$ and remains normalized due to the conservation of momentum. As a result, the excess momentum distribution $\rho_P(i,t)$ assumes the form
\begin{equation}
\rho_P(i,t)=C_P(i,t;j=0,t=0),
\end{equation}
if the kick is put at the atom with index $j=0$.

The centered autocorrelation function of momentum flux $C_{J^P}(t)$ in Eq. (\ref{momentum-diffusion-theory}) is given by
\begin{equation}
C_{J^P}(t)=\frac{1}{N}\langle \Delta J^P(t)\Delta J^P(0)\rangle,\,\,J^P=\sum_i j^P_i,
\end{equation}
where the local momentum flux $j^P_i=-\partial{V(u_i-u_{i-1})}/\partial{u_i}$ with $V(x)$ the form of interaction potential is obtained from the discrete momentum continuity relation
\begin{equation}
\frac{dp_i}{dt}-j^P_i+j^P_{i+1}=0.
\end{equation}

It should be emphasized that here the momentum flux $\Delta J^P(t)$, unlike for energy flux, cannot be replaced with $J^P(t)$ itself. This is so because the equilibrium average is typically non-vanishing with $\langle J^P(t)\rangle=N\Lambda$, where $\Lambda$ denotes a possibly non-vanishing internal equilibrium pressure in cases where the interaction potential is not symmetric.

The transport coefficient of momentum conduction related to the momentum diffusion is the bulk viscosity $\eta$. However, the presence of a finite, isothermal sound speed $v_s$ implies that here the momentum spread contains a ballistic component which should be subtracted~\cite{PhysRev.119.1,Resibois1977} to yield the effective bulk viscosity $\eta$:
\begin{equation}\label{momentum-diffusion-eta}
\eta\equiv\frac{1}{k_BT}\int^{\infty}_{0}C_{J^P}(t)dt-\frac{1}{2}v^2_st.
\end{equation}
In case that the momentum diffusion occurs normal, one can invoke the concept of a finite momentum diffusivity
\begin{equation}\label{momentum-diffusion-dp}
D_P\equiv\frac{1}{2}\lim_{t\rightarrow\infty}\left(\frac{d\langle \Delta x^2(t)\rangle_{\mbox{P}}}{dt}-v^2_st\right).
\end{equation}
Therefore, for the discrete lattices, this effective bulk viscosity $\eta$ precisely equals the momentum diffusivity times the atom mass $m$ (set to unity in the dimensionless units), namely
\begin{equation}
\eta = D_P.
\end{equation}
If the excess momentum density spreads not normally, the limit in Eq. (\ref{momentum-diffusion-dp}) no longer exits. The integration in Eq. (\ref{momentum-diffusion-eta}) formally diverges, thus leading to an infinite effective bulk viscosity.

In the practice, it is found that the finite effective bulk viscosity and normal heat conduction always emerge in pair, so does the infinite effective bulk viscosity and anomalous heat conduction. This constitutes an alternative implementation of the investigation of heat conduction behavior in lattice systems.

\subsubsection{Momentum diffusion properties for typical 1D lattices}

\begin{figure}[t]
\sidecaption
\includegraphics[scale=.5]{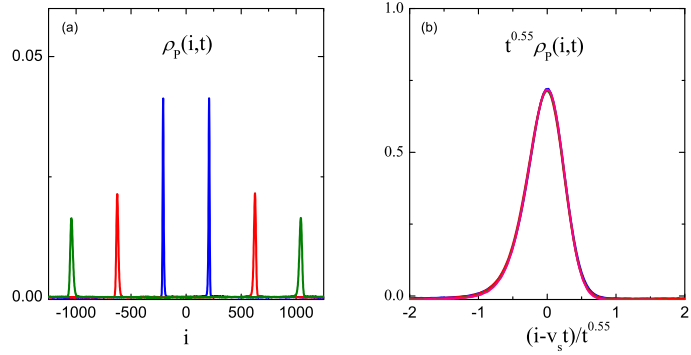}
%
%
\caption{Momentum diffusion processes in the 1D amended coupled rotator lattice. (a) Spatial distribution of the energy autocorrelation $\rho_P(i,t)=C_P(i,t;j=0,t=0)$. The correlation times are $t=200, 600$ and $1000$ from top to the bottom, respectively. There is no central peak and two side peaks moves outside with the constant sound velocity $v_s$. (b) The self-diffusion of the side peak of the momentum spreading. The rescaled side peaks $t^{\delta}\rho_P(i,t)$ of four different correlation times at $t=400,600,800$ and $1000$ all collapse into the same curve in the center of the rescaled moving frame $(i-v_s t)/t^{\delta}$ with a scaling exponent $\delta=0.55$. The scaling exponent $\delta=0.55>0.50$ implies the self-diffusion is superdiffusive and the effective bulk viscosity $\eta$ is infinite. The simulations are performed for an amended coupled rotator lattice with the average energy density per atom $E=1$. The number of atoms $N=2501$ and $K=0.5$.}
\label{fig:Nianbei-amendedrotator-momentum}       
\end{figure}

The momentum diffusion theory connects the momentum diffusion and momentum transport via Eq. (\ref{momentum-diffusion-theory}) and (\ref{momentum-diffusion-eta}). Similar to what we have discussed for the energy diffusion, the excess momentum distribution $\rho_P(i,t)$ is the most important information we need to gather for momentum diffusion method.

We still consider the 1D coupled rotator lattice of Eq. (\ref{cr-ham}) and amended coupled rotator lattice of Eq. (\ref{ar-ham})~\cite{LI_NJP}. In Fig. \ref{fig:Nianbei-rotator-momentum} (a), the excess momentum distributions $\rho_P(i,t)$ for different correlation times $t=200,400$ and $600$ are plotted. At sufficiently large times, the distributions $\rho_P(i,t)$ follow the Gaussian distribution as
\begin{equation}
\rho_P(i,t)\sim\frac{1}{\sqrt{4\pi D_P t}}e^{-\frac{i^2}{4D_P t}},
\end{equation}
where $D_P$ represents the momentum diffusion constant. The MSD $\langle \Delta x^2(t)\rangle_{\mbox{P}}$ of momentum diffusion thus grows linearly with time at large times, i.e. $\langle \Delta x^2(t)\rangle_{\mbox{P}}\sim 2D_P t$, as can be seen from Fig. \ref{fig:Nianbei-rotator-momentum}. There is no ballistic component for the distributions $\rho_P(i,t)$ in 1D coupled rotator lattice. Accordingly, the finite effective bulk viscosity $\eta$ can be simply obtained as $\eta=D_P$. This result is consistent with the fact that the 1D coupled rotator lattice displays normal heat conduction behavior.

For the 1D amended coupled rotator lattice, the energy diffusion is superdiffusive and its heat conduction is anomalous. The excess momentum distributions $\rho_P(i,t)$ for different correlation times at $t=200,600$ and $1000$ are plotted in Fig. \ref{fig:Nianbei-amendedrotator-momentum} (a). In contrast to the coupled rotator lattice, here only two side peaks moving outside with a constant sound velocity $v_s$ exist. To evaluate the true behavior of momentum conduction, this ballistic component within the momentum diffusion should be subtracted. One should instead analyze the self-diffusion behavior of the side peaks of the distributions $\rho_P(i,t)$. In Fig. \ref{fig:Nianbei-amendedrotator-momentum} (b), the rescaled side peaks $t^{\delta}\rho_P(i,t)$ as the function of rescaled position of the peak center $(i-v_s t)/t^{\delta}$ are plotted for four different correlation times at $t=400,600,800$ and $1000$. With the choice of $\delta=0.55$, the rescaled distributions collapse into a single curve all together. This rescaling behavior with $\delta=0.55$ implies a superdiffusive self-diffusion for the side peaks, while normal self-diffusion would require for $\delta=0.50$.

The integration of Eq. (\ref{momentum-diffusion-eta}) is then divergent, giving rise to an infinite effective bulk viscosity $\eta$. This infinite $\eta$ is consistent with the finding that the heat conduction is anomalous, since the energy diffusion is superdiffusive as can be observed in Fig. \ref{fig:Nianbei-amendedrotator-energy} (b). From our perspective and our own numerical results, the infinite bulk viscosity $\eta$ and divergent length-dependent thermal conductivity always emerge in pair. However, there are some other numerical results and approximate theories indicating that finite bulk viscosity and anomalous heat conduction might coexist for symmetric lattices such as FPU-$\beta$ lattice \cite{LeeDadswell2005,Spohn2014jsp}. This is still an open issue and deserves more investigation in the future.

In summary, a novel diffusion method is introduced to investigate the heat transport in 1D nonlinear lattices. The heat and momentum diffusion theories formally relate the diffusion processes to their corresponding conduction processes. The properties of energy and momentum diffusions for typical 1D lattices are presented and more fundamental information about transport processes can be provided from this novel method.

\begin{acknowledgement}
This work is supported by the National Natural Science
Foundation of China under Grant No. 11275267(L.W.), Nos. 11334007 and 11205114 (N.L.),
the Fundamental Research Funds for the Central Universities, and the Research Funds of Renmin University of China(L.W.),
the Program for New Century Excellent Talents of the Ministry of Education of China with Grant No. NCET-12-0409 (N.L.), the Shanghai Rising-Star Program with grant No. 13QA1403600 (N.L.).
Computational resources were provided by the Physical Laboratory of High Performance Computing at Renmin University of China(L.W.) and Shanghai Supercomputer Center (N.L.).
\end{acknowledgement}

\bibliographystyle{spphys}

\end{document}